\documentclass[%
reprint,
amsmath,amssymb,superscriptaddress,
aps,
prx,
]{revtex4-2}

\usepackage{pstricks}
\usepackage{graphicx}
\usepackage{dcolumn}
\usepackage{bm}
\usepackage{color}

\usepackage{physics}

\usepackage{placeins}

\newcommand{\defaultfigurewidth}{0.45\textwidth}
\usepackage{braket}
\usepackage{mathrsfs} 
\usepackage{siunitx}
\usepackage[]{float}

\begin{document}
	
	\preprint{APS/123-QED}

	\title{Neural-network-based design and implementation of fast and robust quantum gates}

	\author{Marko Kuzmanovi\'c}
	\affiliation{InstituteQ and QTF  Centre  of  Excellence,  Department  of  Applied  Physics, School  of  Science,  Aalto  University,  FI-00076  Aalto,  Finland}

	\author{Ilya Moskalenko}
	\affiliation{InstituteQ and QTF  Centre  of  Excellence,  Department  of  Applied  Physics, School  of  Science,  Aalto  University,  FI-00076  Aalto,  Finland}

	\author{Yu-Han Chang}
	\affiliation{InstituteQ and QTF  Centre  of  Excellence,  Department  of  Applied  Physics, School  of  Science,  Aalto  University,  FI-00076  Aalto,  Finland}
	
	\author{Ognjen Stanisavljevi\'c}
	\affiliation{InstituteQ and QTF  Centre  of  Excellence,  Department  of  Applied  Physics, School  of  Science,  Aalto  University,  FI-00076  Aalto,  Finland}
	
	\author{Christopher Warren}
	\affiliation{Department of Microtechnology and Nanoscience (MC2), Chalmers University of Technology, SE-41296 G\"oteborg, Sweden}
	
	\author{Emil Hogedal}
	\affiliation{Department of Microtechnology and Nanoscience (MC2), Chalmers University of Technology, SE-41296 G\"oteborg, Sweden}
	
	\author{Anuj Aggarwal}
	\affiliation{Department of Microtechnology and Nanoscience (MC2), Chalmers University of Technology, SE-41296 G\"oteborg, Sweden}
	
	\author{Irshad Ahmad}
	\affiliation{Department of Microtechnology and Nanoscience (MC2), Chalmers University of Technology, SE-41296 G\"oteborg, Sweden}

	\author{Janka Biznárová}
	\affiliation{Department of Microtechnology and Nanoscience (MC2), Chalmers University of Technology, SE-41296 G\"oteborg, Sweden}
	
	\author{Mamta Dahiya}
	\affiliation{Department of Microtechnology and Nanoscience (MC2), Chalmers University of Technology, SE-41296 G\"oteborg, Sweden}

	\author{Marcus Rommel}
	\affiliation{Department of Microtechnology and Nanoscience (MC2), Chalmers University of Technology, SE-41296 G\"oteborg, Sweden}
	
	\author{Andreas Nylander}
	\affiliation{Department of Microtechnology and Nanoscience (MC2), Chalmers University of Technology, SE-41296 G\"oteborg, Sweden}
	
	\author{Giovanna Tancredi}
	\affiliation{Department of Microtechnology and Nanoscience (MC2), Chalmers University of Technology, SE-41296 G\"oteborg, Sweden}

	\author{Gheorghe Sorin Paraoanu}
	\affiliation{InstituteQ and QTF  Centre  of  Excellence,  Department  of  Applied  Physics, School  of  Science,  Aalto  University,  FI-00076  Aalto,  Finland}

	\date{\today}

	\begin{abstract}
		We present a continuous-time, neural-network-based approach to optimal control in quantum systems, with a focus on pulse engineering for quantum gates. Leveraging the framework of neural ordinary differential equations, we construct control fields as outputs of trainable neural networks, thereby eliminating the need for discrete parametrization or predefined bases. This allows for generation of smooth, hardware-agnostic pulses that can be optimized directly using differentiable integrators.

		As a case study we design, and implement experimentally, a short and detuning-robust $\pi/2$ pulse for photon parity measurements in superconducting transmon circuits. This is achieved through simultaneous optimization for robustness and suppressing the leakage outside of the computational basis. These pulses maintain a fidelity greater than $99.9\%$ over a detuning range of $\approx \pm 20\mathrm{MHz}$, thereby outperforming traditional techniques while retaining comparable gate durations.
		This showcases its potential for high-performance quantum control in experimentally relevant settings.
	\end{abstract}

	\maketitle

	\section{Introduction}

	High-fidelity control of quantum systems is a central requirement across quantum technologies. A key area is quantum computing, where quantum advantage and scalability depend on efficient implementations of quantum error-correction algorithms, which in turn rely on robust, error-limited control of the underlying physical qubits~\cite{Krantz2019,nielsen_chuang_2010}.		

	While standard approaches such as Rabi or DRAG~\cite{motzoi2009} pulses are often sufficient for idealized two/few-level systems with minimal constraints, they fall short in the presence of noise, detuning, and drive amplitude fluctuations. A solution to these problems was first sought through the use of composite pulse (CP) sequences~\cite{cummins2000use}, which do improve robustness but typically suffer from practical limitations, especially when the desired (total) operation is short. Indeed, CP techniques with phased short-pulse sequences maintain high fidelity in broad ranges of parameters but still require many pulses~\cite{Torosov2019,Torosov2022,Harutyunyan2023}; achieving similar performance with a few pulses remains challenging.
	
	More advanced control strategies, such as those based on the GRAPE algorithm~\cite{KHANEJA2005296} or the CRAB method~\cite{caneva2011}, have attempted to address these issues. However, they necessitate either specifying a basis in which the control field is expanded (e.g.,~\cite{PhysRevA.110.062608,caneva2011,kups61563,PhysRevA.106.013107,PhysRevLett.106.190501}), or adopting a discretized approach (e.g.,~\cite{poggi2024universally,Werninghaus2021}). Neural networks (NNs) have also been used in a similar way either by discretizing the control field~\cite{ding2020breaking,ai2021experimentally,Giannelli_2022,Brown_2021} or by way of prescribing a basis~\cite{preti2022continuous} and learning the discrete control parameters. While these approaches generally perform well theoretically, they are associated with certain trade-offs. The discretized signal is not generally smooth and requires large bandwidth for faithful representation due to the discontinuities which are inherently present. These issues can be addressed, as was done in~\cite{Werninghaus2021}, but requires in-situ tuning for a given combination of quantum and control hardware.	Likewise, in addition to the sensitivity of the choice of basis which can be highly problem-dependent~\cite{PhysRevA.110.062608}, higher order terms in the expansion typically also increase the required bandwidth. Additionally, to achieve robustness, these approaches require specific error modeling to achieve first- or second-order cancellation.	
	
	In contrast, neural networks can be leveraged in a manner that simultaneously addresses both sets of problems: NNs can act as universal function approximators~\cite{hornik1989multilayer}, capable of expressing optimal control fields without assuming a fixed basis or discretized representation. The emerging framework of neural ordinary differential equations (Neural ODEs)~\cite{chen2018neuralode}, and the tools developed therein, enables a continuous-time approach whereby a neural network \emph{directly} models the control field as a smooth function of time and is optimized directly through differentiable simulation of the quantum dynamics~\cite{sauvage2022optimal,gungordu2022robust,bhattacharyya2024using}.
	
	In this work, we apply Neural ODEs to the task of designing robust $\pi/2$ pulses for parity measurement in a transmon-based quantum system, which we then verify experimentally. This task requires precise control over qubit rotations under variable detuning, a common situation in cavity QED experiments. Our method produces smooth, hardware-agnostic control fields that outperform standard rectangular and DRAG pulses in robustness while maintaining short durations. This demonstrates that Neural ODEs offer a universal and practical framework for quantum control, capable of addressing challenges that are difficult for traditional approaches. Besides quantum computing,  gates or sequences of gates as considered in this work appear in other applications such as microwave sensing via coherent interaction-free measurements~\cite{McCord2024,McCord2023,Dogra2022} and in protocols for axion detection~\cite{Dixit2021} and magnetometry~\cite{Gusarov2023,Blatter2018,Danilin2018}. Our technique will have immediate applications in continuous-variable quantum computing and quantum memory, enabling the creation of large Schr\"odinger cat states without recalibration of the drive transmon frequency~\cite{PRXQuantum.4.030336}. Additionally, we emphasize the critical role of robust parity mapping operations, implemented via well-calibrated $\pi/2$-pulses that uniformly address a broad frequency range corresponding to highly populated Fock states in the storage cavity. Such precision is essential for both accurate Wigner tomography using Ramsey interferometry with unconditional qubit rotations~\cite{PhysRevLett.89.200402, doi:10.1126/science.1243289} and reliable error syndrome extraction in bosonic cat-qubit architectures~\cite{Sun2014}.

	Sections~\ref{qubit} and~\ref{qutrit} study this problem by considering the qubit and the qutrit subspace, respectively. Section~\ref{methods} gives the details of the neural network approach, and describes the experimental setup used for its verification.

	\section{Methods} \label{methods}
	
	\subsection{Dynamics of a qubit driven by a neural network}
	
	Given a quantum system, its time evolution is governed by the propagator $U$, which satisfies the equation $i \hbar \dot{U}(t) = H(t)U(t)$, with the structure and dimensionality of $U$ and $H$ set by the specifics of the problem under study. In general, the Hamiltonian contains a number of (fixed) parameters describing the system itself and some number of tunable ones, expressing the drive fields that we can manipulate in order to achieve optimal quantum control. These control fields can be represented as complex time-dependent amplitudes that encapsulate the information about the amplitudes and phases of the externally applied control tones.
	For the simplest case of a qubit, this reduces to a single control field $\Omega(t)$, with the Hamiltonian (in the frame co-rotating with the qubit):

	\begin{equation}	
		H = \frac{\hbar}{2}\left[ \Omega(t) \ket{1}\bra{0} + \Omega^*(t) \ket{0}\bra{1} \right].
		\label{eq:hamiltionian}
	\end{equation}		
	
	The control field amplitude $\Omega(t)$ can be identified with the output of a neural network $\Omega(t) = \mathcal{N}(t)$. The equation of motion can now be integrated using a \emph{differentiable} integrator such as "torchdiffeq"~\cite{torchdiffeq}. This allows for the obtained trajectory $U(t)$ to be differentiated with respect to the parameters of the neural network $\mathcal{N}$. Then one can use gradient descent and back-propagation techniques to tune (train) these parameters until a desired trajectory $U(t)$ or final state $U(t_{\mathrm{end}})$ is obtained.
	
	If one wishes to obtain a family of trajectories parametrized by $\theta$ (corresponding to e.g. different rotation angles), the additional parameter can be passed to the neural network $\mathcal{N}(t,\theta)$, such that the entire class of trajectories is optimized for during the training phase, similar to~\cite{sauvage2022optimal,bhattacharyya2024using}. If one wishes to optimize for resilience against deviations in the control parameters, the output of the neural network can be perturbed stochastically in each training iteration, while requiring that the same operation is realized. Unlike previous works~\cite{cummins2000use,gungordu2022robust}, this technique does not require analytical, propagator-level, cancellation of the error terms, allowing for optimization against complex noise and error models. Namely, if one wishes to optimize for amplitude and/or detuning robustness, the control field can be taken as $\Omega(t) = (1 + \alpha) \mathcal{N}(t) e^{i \delta t}$, where the amplitude perturbation $\alpha$ and the detuning perturbation $\delta$ are sampled from appropriately chosen distributions. The training procedure, in the presence of perturbations, is represented schematically in Fig.~\ref{fig:new_vs_old_pulse}. 
	
	In this work we showcase the flexibility of this approach by optimizing the drive pulses for the measurement of photon number parity~\cite{Dixit2021}, by employing a transmon circuit. In short, a photon populates a resonant cavity, which is coupled to a qubit. The interaction Hamiltonian is $H_{\mathrm{int}} = \chi_{\mathrm{disp}} \sigma_z \hat{n}_{\mathrm{ph}}$, resulting in a shift of the qubit transition frequency by $\delta = 2\chi_{\mathrm{disp}}$ per photon in the cavity. Thus the following sequence of operations $\big(\frac{\pi}{2}\big)_X \big(\beta)_Z \big(-\frac{\pi}{2}\big)_X$ (where $\beta = 2\chi_{\mathrm{disp}}\ \langle \hat{n}_{\mathrm{ph}}\rangle \tau$ is the phase accumulated due to the photon associated AC Stark shift during the idle time $\tau = \pi/2\chi_{\mathrm{disp}}$) will result in an identity operation for zero or even number of photons in the detection cavity while it will constitute a $\big(\pi\big)_X$ rotation for one, or the odd-numbered, photon state. For this approach to work the control pulses used to implement the sequence ought to be true $\frac{\pi}{2}$ pulses, even when detuned by $\delta =\hat{n}_{\mathrm{ph}} 2\chi_{\mathrm{disp}}$ from the qubit transition frequency. The problem of engineering \emph{detuning robust} $\frac{\pi}{2}$ pulses is, therefore, used to demonstrate the utility of our neural network approach.
	
	\subsection{Neural network}

	In this work we implement $\mathcal{N}$ as a deep neural network with $\tanh$ activations. The depth $d=3$ and width $w=20$ of the network are chosen as a balance between expressivity and being able to train the network on an ordinary desktop computer in a matter of hours. The network takes two input parameters: the desired rotation angle $\theta$ -- here always set to $\theta = \pi/2$, and the time $t$. The network outputs two real numbers $O_{1,2}$. The first is related to the instantaneous amplitude of the pulse trough $A(t) = \Omega_{\mathrm{max}} \tanh(O_{1})$, while the second one is identified as the pulse phase $\varphi=O_{2}$, resulting in the complex drive field $\Omega(t) = A(t) e^{i \varphi(t)}$. For training purposes, the pulse duration is taken to be $T=2\pi$, with $t \in [-\pi,\pi]$, leading to a natural amplitude scale $\Omega_{2\pi}$ - the amplitude necessary for a $2\pi$ rotation during $T$. In order to limit excessively high pulse amplitudes that might not be feasible experimentally, $\Omega_{\mathrm{max}}$ is set to $\Omega_{\mathrm{max}}= 3\Omega_{2\pi}$.

	For a qubit the propagator $U$ can be parametrized via real coefficients $c_k$ as follows:
	
	\begin{equation}		
		U = c_0 \sigma_0 - i \sum_{k\in{x,y,z}} c_k \sigma_k,
		\label{eq:unitary}
	\end{equation}		
	 with $\sum_{k \in 0,x,y,z} c_k^2 =1$.

	 The E.O.M. then reads $\partial_t\vec{c} = M \vec{c}$, with:
	 \begin{equation}	
	 	M=\frac{1}{2}\left(
	 	\begin{array}{cccc}
	 		0 & -\Re(\Omega ) & -\Im(\Omega) & 0 \\
	 		\Re(\Omega ) & 0 & 0 & \Im(\Omega ) \\
	 		\Im(\Omega ) & 0 & 0 & -\Re(\Omega ) \\
	 		0 & -\Im(\Omega ) & \Re(\Omega ) & 0 \\
	 	\end{array}
	 	\right).
	 	\label{eq:M}
	 \end{equation}
	  
	Since the desired operation is a $\pi/2$ rotation ($c_0 = 1/\sqrt{2}$) in a vertical plane ($c_z=0$, $c_x^2 + c_y^2=1/2$) the loss, or infidelity, of the operation is taken as $L = c_z^2 + (c_0-1/\sqrt{2})^2$. Note that, due to the normalization condition, minimizing this quantity simultaneously ensures that $c_x^2 + c_y^2 \approx 1/2$.
	
\begin{figure}
		\centering
		\includegraphics[width=\defaultfigurewidth]{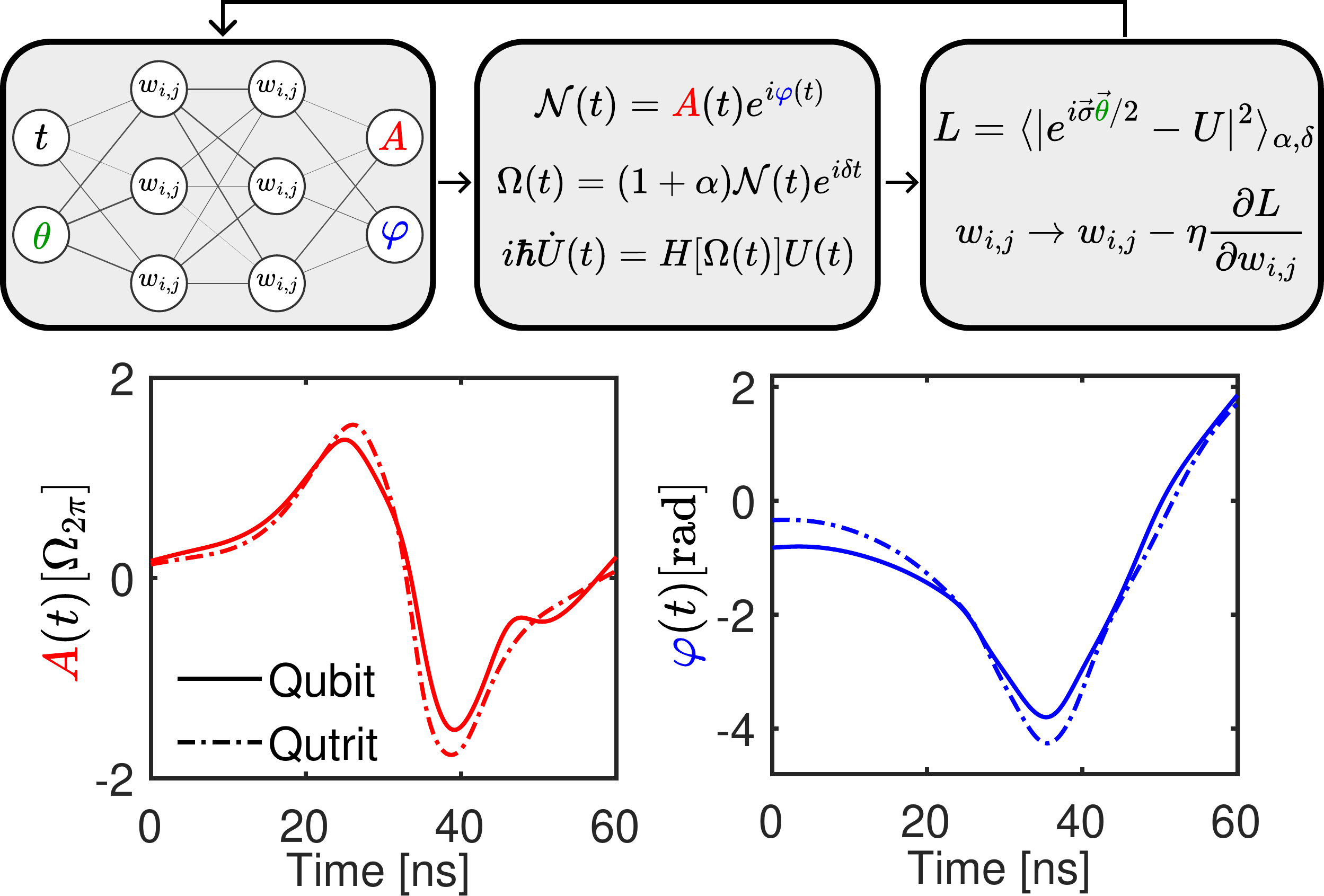}
		\caption{\textbf{Top}: A schematic representation of the Neural network approach: given some target rotation angle $\theta$ and time $t$, the output of the network $\mathcal{N}$ is perturbed by $\alpha$ (amplitude) and $\delta$ (detuning) to construct the complex drive field $\Omega(t)$, and the equation of motion is integrated to obtain the propagator $U$. Based on this a loss $L$ can be computed, which quantifies the discrepancy between the implemented and the target operation. Lastly, the weights of the neural network $w_{i,j}$ are updated based on the gradients of $L$ w.r.t $w_{i,j}$ and a learning rate $\eta$. The cycle is repeated $N$ times or until convergence.			
			\textbf{Bottom}: The resulting pulse amplitude $A(t)$ (left), shown in units of $\Omega_{2\pi} = 2\pi/T$ where $T$ is the pulse duration, and phase $\varphi(t)$ (right) as a function of time, optimized for a true qubit (solid) and a qutrit (dashed-dotted).}
		\label{fig:new_vs_old_pulse}
	\end{figure}

	As amplitude robustness is not considered, $\alpha$ is set to $0$, while $\delta$ is sampled uniformly from the interval $\delta \in [-0.8 \Omega_{2\pi}, 1.1 \Omega_{2\pi}]$. These limits are asymmetric, as only positive detuning is relevant for the parity measurement experiment. While increasing them is feasible, and results in higher detuning robustness, we found that it is achieved by an anomalous pulse whose amplitude is nonzero only close to $t=0$, effectively making it shorter and increasing its bandwidth. In each training iteration, a batch of $100$ to $1000$ $\delta$-values is randomly sampled, and the mean loss is computed. The gradient over all of the neural network parameters is computed using automatic differentiation, and their values are updated by using the "Adam" optimizer~\cite{diederik2014adam}. This whole procedure is repeated until convergence ($\approx 10^4$ iterations). This procedure is summarized in Fig.~\ref{fig:new_vs_old_pulse}.
	
	\subsection{Experimental methods}
	
	The performance of the obtained pulses is verified on a transmon circuit fabricated using the thick-oxide recipe~\cite{PhysRevResearch.5.043001}. The transmon has the transition frequency of $f_{\rm ge} = 3.772 ~\mathrm{GHz}$, anharmonicity $\alpha = - 222.34 ~\mathrm{MHz}$, and relaxation and decoherence times of $T_1=131 ~\mathrm{\mu s}$, $T_{2}=64~\mathrm{\mu s}$, $T_2^{\rm echo}=141~\mathrm{\mu s}$.
	All control signals are generated from the "OPX+" FPGA-based AWG, and up-converted to the microwave domain by the "Octave" up/down-converter module, both manufactured by Quantum Machines. More details about the experimental setup are given in Appendix~\ref{exp_setup}. 
	
	The pulses are characterized through performing standard process tomography~\cite{nielsen_chuang_2010} at different detunings, $\delta$. The resulting data is used to obtain the experimental process matrix $\chi_{\mathrm{exp}}$,  which includes contributions from both finite relaxation times and SPAM errors~\cite{PhysRevLett.109.080505}. The unitary generating said operation is estimated in the following way: for a unitary $U_{\mathrm{fit}} = c^{\mathrm{fit}}_0 \sigma_0 - i \sum_{k\in{x,y,z}} c^{\mathrm{fit}}_k \sigma_k$ the corresponding process matrix $\chi_{\mathrm{fit}}$, in the basis of $\{\sigma_0, \sigma_x, -i\sigma_y, \sigma_z\}$, is given by  $\chi^{j,k}_{\mathrm{fit}} = b_j b^{*}_k$ where 
	$b_{j} = c_{j}$ if $j\in\{0,y\}$ and $b_{j} = -ic_{j}$ if $j\in\{x,z\}$.  We then look for $c^{\mathrm{fit}}_i$ which minimizes the Frobenius norm $|\chi_{\mathrm{exp}} - \chi_{\mathrm{fit}}|^2$, allowing us to extract the underlying operation.
	
	Additionally, the fidelity of the pulses is estimated using a randomized-benchmark inspired~\cite{PhysRevLett.106.180504} sequence, where an "identity" gate is constructed by 4 successive applications of the pulse. This way deviations from a $\pi/2$ rotation angle are observed as oscillations, while the fidelity of the pulse can be extracted from the rate of decay towards the maximally mixed state. Lastly, the same analysis is performed on a rectangular and DRAG~\cite{motzoi2009} $\pi/2$ pulse and their performance is compared.	
		
	\section{Parity measurement}
	\subsection{Qubit}\label{qubit}
	
	With the decomposition for $U$ presented above, the equations of motion can be presented directly in terms of the real and imaginary components of $\Omega$ and the vector of $c_k \in R^4$, allowing for elegant and quick integration.
	Training the neural network results in a pulse like the one presented in Fig.~\ref{fig:new_vs_old_pulse}, shown here in terms of its signed amplitude $A(t)$ and phase $\varphi(t)$. 
	For the performance of the pulse to be competitive with a short DRAG pulse, a pulse duration of $T=60~\mathrm{ns}$ was chosen as a balance between the pulse bandwidth, amplitude, and the width of the robustness region (all of which scale like $\propto T^{-1}$).	To generate the pulse the (logical) $I$ and $Q$ quadratures of the AWG are then set to $I(t) = \Omega(t) \cos(\varphi(t))$ and $Q(t) = \Omega(t) \sin(\varphi(t))$.

	Such modulated pulses can have higher bandwidth and amplitudes resulting in implementation challenges: the finite bandwidth and non-zero nonlinearity of the generation scheme might lead to discrepancies between the desired and the generated pulse. These can be accounted for by measuring the output of the generator with a high-frequency oscilloscope.	Furthermore, the time domain trace can be integrated to estimate the evolution of the qubit under the \textit{generated} drive tone. The results of this procedure, presented in Appendix~\ref{impl-ver}, demonstrate the ability to synthesize these complex pulses.

\begin{figure}
		\centering
		\includegraphics[width=\defaultfigurewidth]{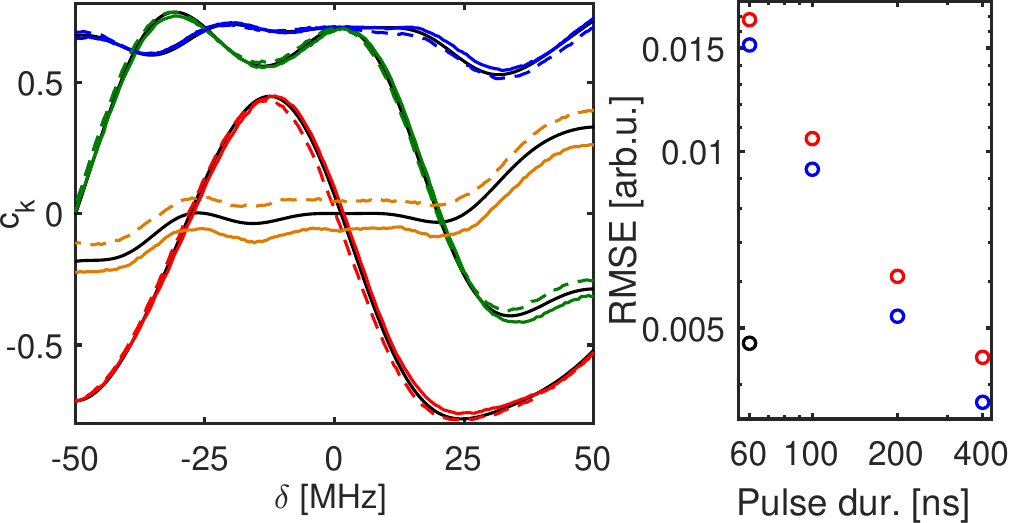}
		\caption{\textbf{Left}: The experimentally obtained coefficients $c_0$ (blue), $c_x$ (red), $c_y$ (green) and $c_z$ (yellow) for the pulse (solid) and its conjugate form (dashed) along with the theoretically expected value (solid black). \textbf{Right}: the RMSE between the theory and the experiment as a function of the pulse duration for the direct (red) and conjugated (blue) pulses, as well as the pulse optimized to suppress the interactions with the $\ket{2}$ state (black).
		}
		\label{fig:old_pulse_proc_tomo}
	\end{figure}

	Fig.~\ref{fig:old_pulse_proc_tomo} shows the results of the process tomography as a function of the detuning frequency. A detailed description of the quantum process tomography procedure can be found in Appendix~\ref{QPT}. The experimental traces (colored) resemble the theory (black), though with some discrepancies. Most notable is an offset of the $c_z$ component w.r.t the expected value. Naively, one might interpret this as a frequency offset between the laboratory frame and the qubit. However the discrepancy remains after careful calibration, and it approximately corresponds to a detuning of $\approx 400\mathrm{kHz}$, well above the level of experimental uncertainty.
	
	Additionally, if one plays the \textit{conjugated} pulse $\Omega(t) =  A(t) e^{-i \varphi(t)}$ we see that the $c_z$ offset changes sign (cf. Fig.~\ref{fig:old_pulse_proc_tomo}), indicating an alternative mechanism. Conjugating the pulse has the same effect as flipping its frequency, or rather detuning with respect to the qubit frequency. Therefore, it also comes with an additional change of sign of $\delta$, which was taken into account for reasons of presentation in Fig.~\ref{fig:old_pulse_proc_tomo}.
	
	The $\varphi(t) \rightarrowtail -\varphi(t)$ and $\delta \rightarrowtail -\delta$ symmetry breaking could have several causes. \textbf{\textit{The first}} is the fact that the signal generation scheme used in this experiment relies on upconversion by mixing the low frequency signal with a local oscillator (LO) close to the qubit frequency. For systems with negative anharmonicity the LO is typically placed above the qubit transition frequency (here we have $f_{\rm LO}=3.9~\mathrm{GHz}$) to avoid driving higher order transitions. However, we find that the LO leakage and image are suppressed by at least $40-50~\mathrm{dBc}$ typically and their contributions are not clearly present in the scope traces. Together with repeating the measurements with different $f_{\rm LO}$ this makes the upconversion scheme an unlikely culprit for the observed effect.
	\textbf{\textit{Another}} possible explanation is that the qubit approximation is not valid, and the symmetry is naturally broken by the presence of higher excited states. To this end, we perform the same type of measurements as shown in Fig.~\ref{fig:old_pulse_proc_tomo} as a function of pulse duration and measure the root-mean-square error (RMSE) between the theory and experiment, for both the non- and conjugated pulses. Fig.~\ref{fig:old_pulse_proc_tomo} shows that the RMSE follows a power law, similar to what is observed when the finite anharmonicity becomes the limiting factor at short pulse durations~\cite{motzoi2009}. The fact that the loss for conjugated pulses is systematically lower than for non-conjugated ones indicates that the two-level approximation is not adequate even for long pulse durations, with low RMSE. It is worthy to note that if the $c_z$ offset was generated by a detuned qubit frequency, or LO leakage, its effect would be exaggerated by increasing the pulse duration.

	\subsection{Qutrit} \label{qutrit}
	
\begin{figure}
		\centering
		\includegraphics[width=\defaultfigurewidth]{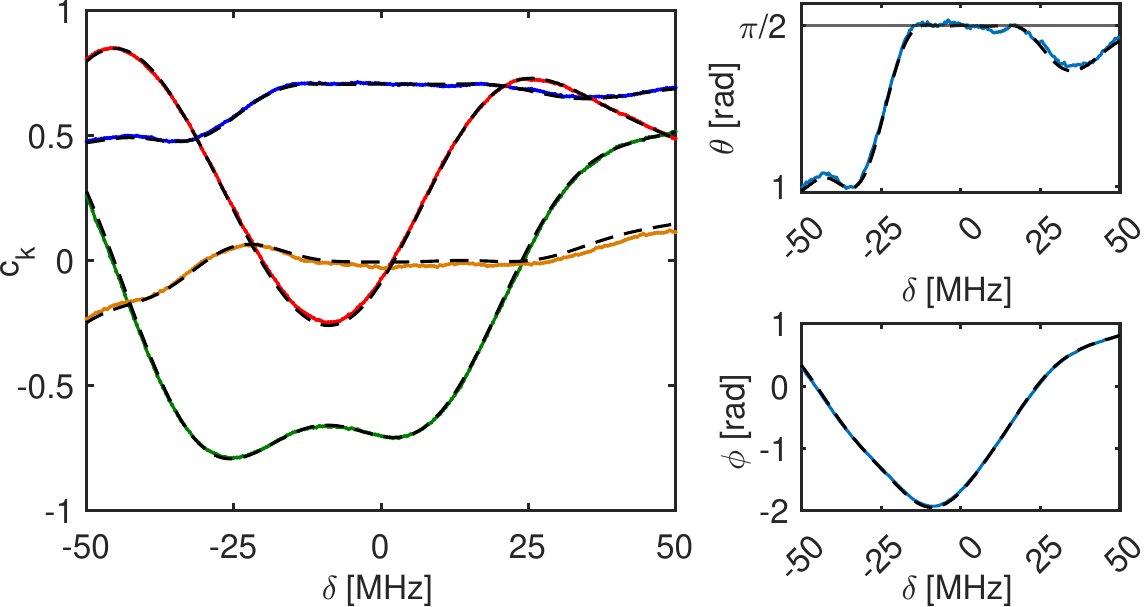}
		\caption{\textbf{Left}: The experimentally obtained coefficients $c_0$ (blue), $c_x$ (red), $c_y$ (green) and $c_z$ (yellow) for the pulse optimized to suppress the interactions with the $\ket{2}$ state. \textbf{Right}: the azimuthal and polar rotation angles, $\theta$ and $\phi$, top and bottom, respectively. The dashed black traces on all panels correspond to the theoretically expected behavior.
		}
		\label{fig:new_pulse_proc_tomo}
	\end{figure}

	With the results obtained in section~\ref{qubit} we turn to optimizing the pulse in the presence of the second excited state, for our specific device: the Hamiltonian of equation~\ref{eq:hamiltionian} is expanded with $\Delta \ket{2}\bra{2}$ and $\lambda\Omega(t)\ket{2}\bra{1} + h.c.$ terms. Here $\Delta = 2\pi \alpha T$ is set by the pulse duration $T$ and the anharmonicity $\alpha$, while $\lambda$ is the ratio of the $\bra{2}\hat{q}\ket{1}$ and $\bra{1}\hat{q}\ket{0}$ matrix elements of the electrical dipole moments. By fitting the transition frequencies of our system, we extract the Josephson and charging energies and evaluate lambda to be $\lambda \approx 1.37$. The $\mathrm{SU(3)}$ group does not permit a linear and real parametrization of the propagator $U$~\cite{bronzan1988parametrization},  forcing the parameter space to be $\mathbb{C}^{3\times3} \cong \mathbb{R}^{18}$, therefore significantly increasing the computational complexity. Thus, we opt to \emph{refine} the pulse, i.e. resume training the neural network from the previous state, thereby reducing the resources needed for the optimization. The resulting pulse is also shown in Fig.~\ref{fig:new_vs_old_pulse}: seemingly minor differences, such as a smoother envelope $A(t)$, lead to a significantly different behavior.	This can be understood by performing QPT as a function of detuning $\delta$, as shown in Fig.~\ref{fig:new_pulse_proc_tomo}. Indeed, we find that $c_0 \approx 1/\sqrt{2}$ and $c_z\approx0$ for a wide range of detuning frequencies, and thus implicitly $c_x^2 + c_y^2=1/2$. When these conditions are met the resulting rotation angle $\theta$ is $\theta \approx \pi/2$, while the plane of rotation $\phi$ is frequency dependent (cf. Fig.~\ref{fig:new_pulse_proc_tomo}). The latter is not an issue for the parity measurement experiment, as the sequence is not sensitive to an overall phase. To quantify the agreement with the theoretically expected behavior the RMSE loss is calculated and shown in Fig.~\ref{fig:old_pulse_proc_tomo} by the black circle: the agreement is significantly improved and is on the level of the error observed for the longest pulse durations.
	
\begin{figure}
		\centering
		\includegraphics[width=\defaultfigurewidth]{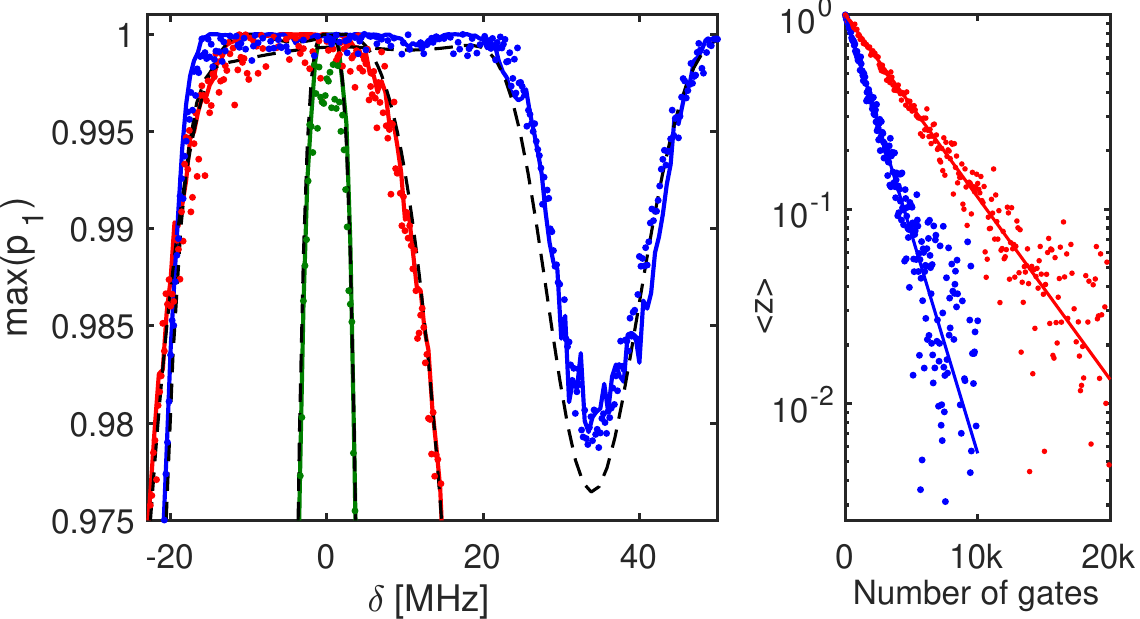}
		\caption{\textbf{Left}: The maximum $\ket{0} \rightarrow \ket{1}$ transfer efficiency as a function of detuning $\delta$ for the robust (blue), DRAG (red) and rectangular (green) pulses. The dots correspond to directly measured values, the solid lines to the QPT inferred ones and the dashed black to the theoretical expectation. \textbf{Right}: The $z$ component of the density as a function of the number of applied robust (blue) and DRAG (red) pulses.}
		\label{fig:p1max_RB}
	\end{figure}

	In the context of the parity measurement experiment the figure of merit can be constructed as the efficiency with which two consequentially applied pulses, with an appropriately chosen delay time in between them, transfer the population of the qubit from the ground $\ket{0}$ state to the excited one $\ket{1}$. With this benchmark in mind, we compare our designer pulse to a rectangular pulse of the same duration $T=60~\mathrm{ns}$ and a $T=20~\mathrm{ns}$ DRAG pulse: we initialize the qubit in the ground state, apply the sequence, and measure the probability to find it in the excited one. The result of this procedure, as a function of the detuning $\delta$ is shown in Fig.~\ref{fig:p1max_RB}.  Here we show the theoretically expected trace along with the experimentally determined values both based on QPT and a direct measurement, for all three pulses. For more details on this procedure see Appendix~\ref{pulse_comparison}.	
	As expected, due to its comparatively small bandwidth, the rectangular pulse performs the worst. The DRAG pulse fares much better, however its efficiency falls off rather quickly for positive detunings $\delta>0$, while it stays high for negative ones: $\max(p_1) >0.99$ for $ -13.5\mathrm{MHz}< \delta  <5\mathrm{MHz}$. This behavior is the result of the \emph{inherent} spectral asymmetry present in the DRAG pulse: the non-zero out-of-phase component $Q(t) \propto \sin(2\pi t/T)$ introduces some phase modulation, thus skewing the spectrum of the pulse. This constitutes an effective detuning, which counteracts the AC Stark shift of the second excited state $\ket{2}$ and cannot be eliminated without degrading the performance of the pulse.

	In contrast, the efficiency of the robust scheme remains high (above $\approx 0.98$) for detunings as large as $\delta = 50~\mathrm{MHz}$, while matching the performance of the DRAG pulse for negative detunings: $\max(p_1) >0.99$ for $ -16.5\mathrm{MHz}< \delta  <22\mathrm{MHz}$ .
	It is worth noting that the sequence implemented by the DRAG and rectangular pulses is not strictly a parity measurement even when the transfer efficiency is high: as the resulting unitary has a non-zero $\sigma_z$ component (cf. Appendix~\ref{pulse_comparison}) the corresponding rotation does not lie in a vertical plane. As a result, the qubit must acquire an additional, photon number dependent, phase to maximize the transfer efficiency. This condition cannot be met simultaneously for all values of $\delta = 2\chi_{\mathrm{disp}} \hat{n}_{\rm ph}$ with a single delay time $\tau$. In contrast, the robust pulse preserves the desired property as $c_z \approx 0$, which is detailed in Appendix~\ref{pulse_comparison} as well.
	
	Lastly, we set out to measure the fidelity of the robust $\frac{\pi}{2}$ pulse and compare it to that of the DRAG. To this end we construct a pseudo-identity operation out of 4 pulses $\big(\frac{\pi}{2}\big)^{4n} = (-1)^n \hat{I}$,
	initialize the system in the ground state and measure the $z$ component of the density matrix as a function of the number of repetitions $n$. 	We perform this procedure only on resonance ($\delta=0$) - see Fig.~\ref{fig:p1max_RB}. By fitting the results with an exponential decay we are able to extract the fidelity per pulse: $F_{\rm rob} = 0.999481 \pm 0.000017$ and $F_{\rm DRAG} = 0.999784 \pm 0.000008$. 	
	
	The estimation for the single-qubit incoherent errors corresponding to a gate length $T$ with relaxation $T_1$ and pure Markovian dephasing $T_{\phi}$ is given by~\cite{PhysRevX.13.031035,abad2022universal}: $F^{\mathrm{max}}=1-T/3(1/T_1 + 1/T_{\phi} )$, where $T_{\phi}$ is estimated as $1/T_{\phi}=1/T_{2} - 1/2T_{1}$. For our experimental parameters this yields
	$F^{\mathrm{max}}_{\rm 60ns}=0.999611$ and $F^{\mathrm{max}}_{\rm 20ns}=0.999870$. Therefore, the total incoherent error constitutes approximately $75\%$ of the total robust pulse error and $60\%$ of the total DRAG pulse error. The remaining $25\%$ and $40\%$ we assign to the gate implementation and leakage errors, demonstrating a better implementation fidelity of the robust pulse compared to the DRAG. As a result we can conclude that the performance of the pulses is limited by finite relaxation time of the transmon and not by the implementation.

	\bigskip
	
	\section{Discussion and conclusions} \label{discussion}

	We presented a neural ordinary differential equation (Neural ODE) framework for quantum control, enabling continuous-time, basis-free synthesis of high-fidelity control pulses. 
	By representing drive fields as outputs of trainable neural networks, and optimizing them through automatic differentiation, our approach provides universal, hardware-agnostic control strategies without relying on predefined bases or discretized control points. 
	
	We validate this method by generating robust $\pi/2$ pulses for parity measurements in superconducting qubits, achieving superior detuning resilience compared to standard rectangular and DRAG pulses. This work positions Neural ODEs as a powerful tool for general-purpose quantum gate optimization.

	\vspace{3mm}
	
	\section*{\bf AUTHOR CONTRIBUTIONS} 
	
	MK carried out the modeling, numerical optimization and together with IM, YHC, and OS performed the experiments. The Chalmers team designed and fabricated the transmon device. All members of the Aalto team contributed to the writing of the manuscript. GSP supervised the project.

	\acknowledgments	
	
	We thank Miroslav Bogdanovi\'c for assistance with the machine-learning portion of the work, and Jonas Bylander for comments on the manuscript. This project has received funding from the EU Flagship on Quantum Technology through HORIZON-CL4-2022-QUANTUM-01-SGA Project No. 101113946 OpenSuperQPlus100.\;\; We also acknowledge support from Research Council of Finland Proof-of-Concept program (Project 359397) and from the Centre of Excellence program (Project 352925). The  Chalmers team acknowledges
	financial support by the Knut and Alice Wallenberg
	through the Wallenberg Center for Quantum Technology
	(WACQT). The fabrication of the quantum processor was performed at Myfab Chalmers. The Aalto team acknowledges the provision of facilities and technical support by the Aalto University at the national research infrastructure OtaNano - Low Temperature Laboratory.

	\FloatBarrier
	
	\bibliography{Bibliography_NN}

\providecommand{\noopsort}[1]{}\providecommand{\singleletter}[1]{#1}%
\begin{thebibliography}{50}%
\makeatletter
\providecommand \@ifxundefined [1]{%
 \@ifx{#1\undefined}
}%
\providecommand \@ifnum [1]{%
 \ifnum #1\expandafter \@firstoftwo
 \else \expandafter \@secondoftwo
 \fi
}%
\providecommand \@ifx [1]{%
 \ifx #1\expandafter \@firstoftwo
 \else \expandafter \@secondoftwo
 \fi
}%
\providecommand \natexlab [1]{#1}%
\providecommand \enquote  [1]{``#1''}%
\providecommand \bibnamefont  [1]{#1}%
\providecommand \bibfnamefont [1]{#1}%
\providecommand \citenamefont [1]{#1}%
\providecommand \href@noop [0]{\@secondoftwo}%
\providecommand \href [0]{\begingroup \@sanitize@url \@href}%
\providecommand \@href[1]{\@@startlink{#1}\@@href}%
\providecommand \@@href[1]{\endgroup#1\@@endlink}%
\providecommand \@sanitize@url [0]{\catcode `\\12\catcode `\$12\catcode
  `\&12\catcode `\#12\catcode `\^12\catcode `\_12\catcode `\%12\relax}%
\providecommand \@@startlink[1]{}%
\providecommand \@@endlink[0]{}%
\providecommand \url  [0]{\begingroup\@sanitize@url \@url }%
\providecommand \@url [1]{\endgroup\@href {#1}{\urlprefix }}%
\providecommand \urlprefix  [0]{URL }%
\providecommand \Eprint [0]{\href }%
\providecommand \doibase [0]{https://doi.org/}%
\providecommand \selectlanguage [0]{\@gobble}%
\providecommand \bibinfo  [0]{\@secondoftwo}%
\providecommand \bibfield  [0]{\@secondoftwo}%
\providecommand \translation [1]{[#1]}%
\providecommand \BibitemOpen [0]{}%
\providecommand \bibitemStop [0]{}%
\providecommand \bibitemNoStop [0]{.\EOS\space}%
\providecommand \EOS [0]{\spacefactor3000\relax}%
\providecommand \BibitemShut  [1]{\csname bibitem#1\endcsname}%
\let\auto@bib@innerbib\@empty
\bibitem [{\citenamefont {Krantz}\ \emph {et~al.}(2019)\citenamefont {Krantz},
  \citenamefont {Kjaergaard}, \citenamefont {Yan}, \citenamefont {Orlando},
  \citenamefont {Gustavsson},\ and\ \citenamefont {Oliver}}]{Krantz2019}%
  \BibitemOpen
  \bibfield  {author} {\bibinfo {author} {\bibfnamefont {P.}~\bibnamefont
  {Krantz}}, \bibinfo {author} {\bibfnamefont {M.}~\bibnamefont {Kjaergaard}},
  \bibinfo {author} {\bibfnamefont {F.}~\bibnamefont {Yan}}, \bibinfo {author}
  {\bibfnamefont {T.~P.}\ \bibnamefont {Orlando}}, \bibinfo {author}
  {\bibfnamefont {S.}~\bibnamefont {Gustavsson}},\ and\ \bibinfo {author}
  {\bibfnamefont {W.~D.}\ \bibnamefont {Oliver}},\ }\bibfield  {title}
  {\bibinfo {title} {A quantum engineer's guide to superconducting qubits},\
  }\href@noop {} {\bibfield  {journal} {\bibinfo  {journal} {Applied physics
  reviews}\ }\textbf {\bibinfo {volume} {6}} (\bibinfo {year}
  {2019})}\BibitemShut {NoStop}%
\bibitem [{\citenamefont {Nielsen}\ and\ \citenamefont
  {Chuang}(2010)}]{nielsen_chuang_2010}%
  \BibitemOpen
  \bibfield  {author} {\bibinfo {author} {\bibfnamefont {M.~A.}\ \bibnamefont
  {Nielsen}}\ and\ \bibinfo {author} {\bibfnamefont {I.~L.}\ \bibnamefont
  {Chuang}},\ }\href {https://doi.org/10.1017/CBO9780511976667} {\emph
  {\bibinfo {title} {Quantum Computation and Quantum Information: 10th
  Anniversary Edition}}}\ (\bibinfo  {publisher} {Cambridge University Press},\
  \bibinfo {year} {2010})\BibitemShut {NoStop}%
\bibitem [{\citenamefont {Motzoi}\ \emph {et~al.}(2009)\citenamefont {Motzoi},
  \citenamefont {Gambetta}, \citenamefont {Rebentrost},\ and\ \citenamefont
  {Wilhelm}}]{motzoi2009}%
  \BibitemOpen
  \bibfield  {author} {\bibinfo {author} {\bibfnamefont {F.}~\bibnamefont
  {Motzoi}}, \bibinfo {author} {\bibfnamefont {J.~M.}\ \bibnamefont
  {Gambetta}}, \bibinfo {author} {\bibfnamefont {P.}~\bibnamefont
  {Rebentrost}},\ and\ \bibinfo {author} {\bibfnamefont {F.~K.}\ \bibnamefont
  {Wilhelm}},\ }\bibfield  {title} {\bibinfo {title} {Simple pulses for
  elimination of leakage in weakly nonlinear qubits},\ }\href@noop {}
  {\bibfield  {journal} {\bibinfo  {journal} {Physical Review Letters}\
  }\textbf {\bibinfo {volume} {103}},\ \bibinfo {pages} {110501} (\bibinfo
  {year} {2009})}\BibitemShut {NoStop}%
\bibitem [{\citenamefont {Cummins}\ and\ \citenamefont
  {Jones}(2000)}]{cummins2000use}%
  \BibitemOpen
  \bibfield  {author} {\bibinfo {author} {\bibfnamefont {H.~K.}\ \bibnamefont
  {Cummins}}\ and\ \bibinfo {author} {\bibfnamefont {J.}~\bibnamefont
  {Jones}},\ }\bibfield  {title} {\bibinfo {title} {Use of composite rotations
  to correct systematic errors in nmr quantum computation},\ }\href@noop {}
  {\bibfield  {journal} {\bibinfo  {journal} {New Journal of Physics}\ }\textbf
  {\bibinfo {volume} {2}},\ \bibinfo {pages} {6} (\bibinfo {year}
  {2000})}\BibitemShut {NoStop}%
\bibitem [{\citenamefont {Torosov}\ and\ \citenamefont
  {Vitanov}(2019)}]{Torosov2019}%
  \BibitemOpen
  \bibfield  {author} {\bibinfo {author} {\bibfnamefont {B.~T.}\ \bibnamefont
  {Torosov}}\ and\ \bibinfo {author} {\bibfnamefont {N.~V.}\ \bibnamefont
  {Vitanov}},\ }\bibfield  {title} {\bibinfo {title} {Arbitrarily accurate
  variable rotations on the bloch sphere by composite pulse sequences},\ }\href
  {https://doi.org/10.1103/PhysRevA.99.013402} {\bibfield  {journal} {\bibinfo
  {journal} {Phys. Rev. A}\ }\textbf {\bibinfo {volume} {99}},\ \bibinfo
  {pages} {013402} (\bibinfo {year} {2019})}\BibitemShut {NoStop}%
\bibitem [{\citenamefont {Torosov}\ and\ \citenamefont
  {Vitanov}(2022)}]{Torosov2022}%
  \BibitemOpen
  \bibfield  {author} {\bibinfo {author} {\bibfnamefont {B.~T.}\ \bibnamefont
  {Torosov}}\ and\ \bibinfo {author} {\bibfnamefont {N.~V.}\ \bibnamefont
  {Vitanov}},\ }\bibfield  {title} {\bibinfo {title} {Experimental
  demonstration of composite pulses on ibm's quantum computer},\ }\href
  {https://doi.org/10.1103/PhysRevApplied.18.034062} {\bibfield  {journal}
  {\bibinfo  {journal} {Phys. Rev. Appl.}\ }\textbf {\bibinfo {volume} {18}},\
  \bibinfo {pages} {034062} (\bibinfo {year} {2022})}\BibitemShut {NoStop}%
\bibitem [{\citenamefont {Harutyunyan}\ \emph {et~al.}(2023)\citenamefont
  {Harutyunyan}, \citenamefont {Holweck}, \citenamefont {Sugny},\ and\
  \citenamefont {Gu\'erin}}]{Harutyunyan2023}%
  \BibitemOpen
  \bibfield  {author} {\bibinfo {author} {\bibfnamefont {M.}~\bibnamefont
  {Harutyunyan}}, \bibinfo {author} {\bibfnamefont {F.}~\bibnamefont
  {Holweck}}, \bibinfo {author} {\bibfnamefont {D.}~\bibnamefont {Sugny}},\
  and\ \bibinfo {author} {\bibfnamefont {S.}~\bibnamefont {Gu\'erin}},\
  }\bibfield  {title} {\bibinfo {title} {Digital optimal robust control},\
  }\href {https://doi.org/10.1103/PhysRevLett.131.200801} {\bibfield  {journal}
  {\bibinfo  {journal} {Phys. Rev. Lett.}\ }\textbf {\bibinfo {volume} {131}},\
  \bibinfo {pages} {200801} (\bibinfo {year} {2023})}\BibitemShut {NoStop}%
\bibitem [{\citenamefont {Khaneja}\ \emph {et~al.}(2005)\citenamefont
  {Khaneja}, \citenamefont {Reiss}, \citenamefont {Kehlet}, \citenamefont
  {Schulte-Herbrüggen},\ and\ \citenamefont {Glaser}}]{KHANEJA2005296}%
  \BibitemOpen
  \bibfield  {author} {\bibinfo {author} {\bibfnamefont {N.}~\bibnamefont
  {Khaneja}}, \bibinfo {author} {\bibfnamefont {T.}~\bibnamefont {Reiss}},
  \bibinfo {author} {\bibfnamefont {C.}~\bibnamefont {Kehlet}}, \bibinfo
  {author} {\bibfnamefont {T.}~\bibnamefont {Schulte-Herbrüggen}},\ and\
  \bibinfo {author} {\bibfnamefont {S.~J.}\ \bibnamefont {Glaser}},\ }\bibfield
   {title} {\bibinfo {title} {Optimal control of coupled spin dynamics: design
  of nmr pulse sequences by gradient ascent algorithms},\ }\href
  {https://doi.org/https://doi.org/10.1016/j.jmr.2004.11.004} {\bibfield
  {journal} {\bibinfo  {journal} {Journal of Magnetic Resonance}\ }\textbf
  {\bibinfo {volume} {172}},\ \bibinfo {pages} {296} (\bibinfo {year}
  {2005})}\BibitemShut {NoStop}%
\bibitem [{\citenamefont {Caneva}\ \emph {et~al.}(2011)\citenamefont {Caneva},
  \citenamefont {Calarco},\ and\ \citenamefont {Montangero}}]{caneva2011}%
  \BibitemOpen
  \bibfield  {author} {\bibinfo {author} {\bibfnamefont {T.}~\bibnamefont
  {Caneva}}, \bibinfo {author} {\bibfnamefont {T.}~\bibnamefont {Calarco}},\
  and\ \bibinfo {author} {\bibfnamefont {S.}~\bibnamefont {Montangero}},\
  }\bibfield  {title} {\bibinfo {title} {Chopped random-basis quantum
  optimization},\ }\href@noop {} {\bibfield  {journal} {\bibinfo  {journal}
  {Physical Review A}\ }\textbf {\bibinfo {volume} {84}},\ \bibinfo {pages}
  {022326} (\bibinfo {year} {2011})}\BibitemShut {NoStop}%
\bibitem [{\citenamefont {Pagano}\ \emph {et~al.}(2024)\citenamefont {Pagano},
  \citenamefont {M\"uller}, \citenamefont {Calarco}, \citenamefont
  {Montangero},\ and\ \citenamefont {Rembold}}]{PhysRevA.110.062608}%
  \BibitemOpen
  \bibfield  {author} {\bibinfo {author} {\bibfnamefont {A.}~\bibnamefont
  {Pagano}}, \bibinfo {author} {\bibfnamefont {M.~M.}\ \bibnamefont
  {M\"uller}}, \bibinfo {author} {\bibfnamefont {T.}~\bibnamefont {Calarco}},
  \bibinfo {author} {\bibfnamefont {S.}~\bibnamefont {Montangero}},\ and\
  \bibinfo {author} {\bibfnamefont {P.}~\bibnamefont {Rembold}},\ }\bibfield
  {title} {\bibinfo {title} {Role of bases in quantum optimal control},\ }\href
  {https://doi.org/10.1103/PhysRevA.110.062608} {\bibfield  {journal} {\bibinfo
   {journal} {Phys. Rev. A}\ }\textbf {\bibinfo {volume} {110}},\ \bibinfo
  {pages} {062608} (\bibinfo {year} {2024})}\BibitemShut {NoStop}%
\bibitem [{\citenamefont {Rembold}(2022)}]{kups61563}%
  \BibitemOpen
  \bibfield  {author} {\bibinfo {author} {\bibfnamefont {P.}~\bibnamefont
  {Rembold}},\ }\emph {\bibinfo {title} {Quantum Optimal Control of Spin
  Systems and Trapped Atoms}},\ \href {https://kups.ub.uni-koeln.de/61563/}
  {Ph.D. thesis},\ \bibinfo  {school} {Universit{\"a}t zu K{\"o}ln} (\bibinfo
  {year} {2022})\BibitemShut {NoStop}%
\bibitem [{\citenamefont {Oshnik}\ \emph {et~al.}(2022)\citenamefont {Oshnik},
  \citenamefont {Rembold}, \citenamefont {Calarco}, \citenamefont {Montangero},
  \citenamefont {Neu},\ and\ \citenamefont {M\"uller}}]{PhysRevA.106.013107}%
  \BibitemOpen
  \bibfield  {author} {\bibinfo {author} {\bibfnamefont {N.}~\bibnamefont
  {Oshnik}}, \bibinfo {author} {\bibfnamefont {P.}~\bibnamefont {Rembold}},
  \bibinfo {author} {\bibfnamefont {T.}~\bibnamefont {Calarco}}, \bibinfo
  {author} {\bibfnamefont {S.}~\bibnamefont {Montangero}}, \bibinfo {author}
  {\bibfnamefont {E.}~\bibnamefont {Neu}},\ and\ \bibinfo {author}
  {\bibfnamefont {M.~M.}\ \bibnamefont {M\"uller}},\ }\bibfield  {title}
  {\bibinfo {title} {Robust magnetometry with single nitrogen-vacancy centers
  via two-step optimization},\ }\href
  {https://doi.org/10.1103/PhysRevA.106.013107} {\bibfield  {journal} {\bibinfo
   {journal} {Phys. Rev. A}\ }\textbf {\bibinfo {volume} {106}},\ \bibinfo
  {pages} {013107} (\bibinfo {year} {2022})}\BibitemShut {NoStop}%
\bibitem [{\citenamefont {Doria}\ \emph {et~al.}(2011)\citenamefont {Doria},
  \citenamefont {Calarco},\ and\ \citenamefont
  {Montangero}}]{PhysRevLett.106.190501}%
  \BibitemOpen
  \bibfield  {author} {\bibinfo {author} {\bibfnamefont {P.}~\bibnamefont
  {Doria}}, \bibinfo {author} {\bibfnamefont {T.}~\bibnamefont {Calarco}},\
  and\ \bibinfo {author} {\bibfnamefont {S.}~\bibnamefont {Montangero}},\
  }\bibfield  {title} {\bibinfo {title} {Optimal control technique for
  many-body quantum dynamics},\ }\href
  {https://doi.org/10.1103/PhysRevLett.106.190501} {\bibfield  {journal}
  {\bibinfo  {journal} {Phys. Rev. Lett.}\ }\textbf {\bibinfo {volume} {106}},\
  \bibinfo {pages} {190501} (\bibinfo {year} {2011})}\BibitemShut {NoStop}%
\bibitem [{\citenamefont {Poggi}\ \emph {et~al.}(2024)\citenamefont {Poggi},
  \citenamefont {De~Chiara}, \citenamefont {Campbell},\ and\ \citenamefont
  {Kiely}}]{poggi2024universally}%
  \BibitemOpen
  \bibfield  {author} {\bibinfo {author} {\bibfnamefont {P.~M.}\ \bibnamefont
  {Poggi}}, \bibinfo {author} {\bibfnamefont {G.}~\bibnamefont {De~Chiara}},
  \bibinfo {author} {\bibfnamefont {S.}~\bibnamefont {Campbell}},\ and\
  \bibinfo {author} {\bibfnamefont {A.}~\bibnamefont {Kiely}},\ }\bibfield
  {title} {\bibinfo {title} {Universally robust quantum control},\ }\href@noop
  {} {\bibfield  {journal} {\bibinfo  {journal} {Physical review letters}\
  }\textbf {\bibinfo {volume} {132}},\ \bibinfo {pages} {193801} (\bibinfo
  {year} {2024})}\BibitemShut {NoStop}%
\bibitem [{\citenamefont {Werninghaus}\ \emph {et~al.}(2021)\citenamefont
  {Werninghaus}, \citenamefont {Egger}, \citenamefont {Roy}, \citenamefont
  {Machnes}, \citenamefont {Wilhelm},\ and\ \citenamefont
  {Filipp}}]{Werninghaus2021}%
  \BibitemOpen
  \bibfield  {author} {\bibinfo {author} {\bibfnamefont {M.}~\bibnamefont
  {Werninghaus}}, \bibinfo {author} {\bibfnamefont {D.~J.}\ \bibnamefont
  {Egger}}, \bibinfo {author} {\bibfnamefont {F.}~\bibnamefont {Roy}}, \bibinfo
  {author} {\bibfnamefont {S.}~\bibnamefont {Machnes}}, \bibinfo {author}
  {\bibfnamefont {F.~K.}\ \bibnamefont {Wilhelm}},\ and\ \bibinfo {author}
  {\bibfnamefont {S.}~\bibnamefont {Filipp}},\ }\bibfield  {title} {\bibinfo
  {title} {Leakage reduction in fast superconducting qubit gates via optimal
  control},\ }\href {https://doi.org/10.1038/s41534-020-00346-2} {\bibfield
  {journal} {\bibinfo  {journal} {npj Quantum Information}\ }\textbf {\bibinfo
  {volume} {7}},\ \bibinfo {pages} {14} (\bibinfo {year} {2021})}\BibitemShut
  {NoStop}%
\bibitem [{\citenamefont {Ding}\ \emph {et~al.}(2021)\citenamefont {Ding},
  \citenamefont {Ban}, \citenamefont {Mart{\'\i}n-Guerrero}, \citenamefont
  {Solano}, \citenamefont {Casanova},\ and\ \citenamefont
  {Chen}}]{ding2020breaking}%
  \BibitemOpen
  \bibfield  {author} {\bibinfo {author} {\bibfnamefont {Y.}~\bibnamefont
  {Ding}}, \bibinfo {author} {\bibfnamefont {Y.}~\bibnamefont {Ban}}, \bibinfo
  {author} {\bibfnamefont {J.~D.}\ \bibnamefont {Mart{\'\i}n-Guerrero}},
  \bibinfo {author} {\bibfnamefont {E.}~\bibnamefont {Solano}}, \bibinfo
  {author} {\bibfnamefont {J.}~\bibnamefont {Casanova}},\ and\ \bibinfo
  {author} {\bibfnamefont {X.}~\bibnamefont {Chen}},\ }\bibfield  {title}
  {\bibinfo {title} {Breaking adiabatic quantum control with deep learning},\
  }\href@noop {} {\bibfield  {journal} {\bibinfo  {journal} {Physical Review
  A}\ }\textbf {\bibinfo {volume} {103}},\ \bibinfo {pages} {L040401} (\bibinfo
  {year} {2021})}\BibitemShut {NoStop}%
\bibitem [{\citenamefont {Ai}\ \emph {et~al.}(2022)\citenamefont {Ai},
  \citenamefont {Ding}, \citenamefont {Ban}, \citenamefont
  {Mart{\'\i}n-Guerrero}, \citenamefont {Casanova}, \citenamefont {Cui},
  \citenamefont {Huang}, \citenamefont {Chen}, \citenamefont {Li},\ and\
  \citenamefont {Guo}}]{ai2021experimentally}%
  \BibitemOpen
  \bibfield  {author} {\bibinfo {author} {\bibfnamefont {M.-Z.}\ \bibnamefont
  {Ai}}, \bibinfo {author} {\bibfnamefont {Y.}~\bibnamefont {Ding}}, \bibinfo
  {author} {\bibfnamefont {Y.}~\bibnamefont {Ban}}, \bibinfo {author}
  {\bibfnamefont {J.~D.}\ \bibnamefont {Mart{\'\i}n-Guerrero}}, \bibinfo
  {author} {\bibfnamefont {J.}~\bibnamefont {Casanova}}, \bibinfo {author}
  {\bibfnamefont {J.-M.}\ \bibnamefont {Cui}}, \bibinfo {author} {\bibfnamefont
  {Y.-F.}\ \bibnamefont {Huang}}, \bibinfo {author} {\bibfnamefont
  {X.}~\bibnamefont {Chen}}, \bibinfo {author} {\bibfnamefont {C.-F.}\
  \bibnamefont {Li}},\ and\ \bibinfo {author} {\bibfnamefont {G.-C.}\
  \bibnamefont {Guo}},\ }\bibfield  {title} {\bibinfo {title} {Experimentally
  realizing efficient quantum control with reinforcement learning},\ }\href
  {https://doi.org/10.1007/s11433-021-1841-2} {\bibfield  {journal} {\bibinfo
  {journal} {Sci. China Phys. Mech. Astron.}\ }\textbf {\bibinfo {volume}
  {65}},\ \bibinfo {pages} {250312} (\bibinfo {year} {2022})}\BibitemShut
  {NoStop}%
\bibitem [{\citenamefont {Giannelli}\ \emph {et~al.}(2022)\citenamefont
  {Giannelli}, \citenamefont {Sgroi}, \citenamefont {Brown}, \citenamefont
  {Paraoanu}, \citenamefont {Paternostro}, \citenamefont {Paladino},\ and\
  \citenamefont {Falci}}]{Giannelli_2022}%
  \BibitemOpen
  \bibfield  {author} {\bibinfo {author} {\bibfnamefont {L.}~\bibnamefont
  {Giannelli}}, \bibinfo {author} {\bibfnamefont {P.}~\bibnamefont {Sgroi}},
  \bibinfo {author} {\bibfnamefont {J.}~\bibnamefont {Brown}}, \bibinfo
  {author} {\bibfnamefont {G.~S.}\ \bibnamefont {Paraoanu}}, \bibinfo {author}
  {\bibfnamefont {M.}~\bibnamefont {Paternostro}}, \bibinfo {author}
  {\bibfnamefont {E.}~\bibnamefont {Paladino}},\ and\ \bibinfo {author}
  {\bibfnamefont {G.}~\bibnamefont {Falci}},\ }\bibfield  {title} {\bibinfo
  {title} {A tutorial on optimal control and reinforcement learning methods for
  quantum technologies},\ }\href
  {https://doi.org/https://doi.org/10.1016/j.physleta.2022.128054} {\bibfield
  {journal} {\bibinfo  {journal} {Physics Letters A}\ }\textbf {\bibinfo
  {volume} {434}},\ \bibinfo {pages} {128054} (\bibinfo {year}
  {2022})}\BibitemShut {NoStop}%
\bibitem [{\citenamefont {Brown}\ \emph {et~al.}(2021)\citenamefont {Brown},
  \citenamefont {Sgroi}, \citenamefont {Giannelli}, \citenamefont {Paraoanu},
  \citenamefont {Paladino}, \citenamefont {Falci}, \citenamefont
  {Paternostro},\ and\ \citenamefont {Ferraro}}]{Brown_2021}%
  \BibitemOpen
  \bibfield  {author} {\bibinfo {author} {\bibfnamefont {J.}~\bibnamefont
  {Brown}}, \bibinfo {author} {\bibfnamefont {P.}~\bibnamefont {Sgroi}},
  \bibinfo {author} {\bibfnamefont {L.}~\bibnamefont {Giannelli}}, \bibinfo
  {author} {\bibfnamefont {G.~S.}\ \bibnamefont {Paraoanu}}, \bibinfo {author}
  {\bibfnamefont {E.}~\bibnamefont {Paladino}}, \bibinfo {author}
  {\bibfnamefont {G.}~\bibnamefont {Falci}}, \bibinfo {author} {\bibfnamefont
  {M.}~\bibnamefont {Paternostro}},\ and\ \bibinfo {author} {\bibfnamefont
  {A.}~\bibnamefont {Ferraro}},\ }\bibfield  {title} {\bibinfo {title}
  {Reinforcement learning-enhanced protocols for coherent population-transfer
  in three-level quantum systems},\ }\href
  {https://doi.org/10.1088/1367-2630/ac2393} {\bibfield  {journal} {\bibinfo
  {journal} {New Journal of Physics}\ }\textbf {\bibinfo {volume} {23}},\
  \bibinfo {pages} {093035} (\bibinfo {year} {2021})}\BibitemShut {NoStop}%
\bibitem [{\citenamefont {Preti}\ \emph {et~al.}(2022)\citenamefont {Preti},
  \citenamefont {Calarco},\ and\ \citenamefont {Motzoi}}]{preti2022continuous}%
  \BibitemOpen
  \bibfield  {author} {\bibinfo {author} {\bibfnamefont {F.}~\bibnamefont
  {Preti}}, \bibinfo {author} {\bibfnamefont {T.}~\bibnamefont {Calarco}},\
  and\ \bibinfo {author} {\bibfnamefont {F.}~\bibnamefont {Motzoi}},\
  }\bibfield  {title} {\bibinfo {title} {Continuous quantum gate sets and
  pulse-class meta-optimization},\ }\href@noop {} {\bibfield  {journal}
  {\bibinfo  {journal} {PRX Quantum}\ }\textbf {\bibinfo {volume} {3}},\
  \bibinfo {pages} {040311} (\bibinfo {year} {2022})}\BibitemShut {NoStop}%
\bibitem [{\citenamefont {Hornik}\ \emph {et~al.}(1989)\citenamefont {Hornik},
  \citenamefont {Stinchcombe},\ and\ \citenamefont
  {White}}]{hornik1989multilayer}%
  \BibitemOpen
  \bibfield  {author} {\bibinfo {author} {\bibfnamefont {K.}~\bibnamefont
  {Hornik}}, \bibinfo {author} {\bibfnamefont {M.}~\bibnamefont
  {Stinchcombe}},\ and\ \bibinfo {author} {\bibfnamefont {H.}~\bibnamefont
  {White}},\ }\bibfield  {title} {\bibinfo {title} {Multilayer feedforward
  networks are universal approximators},\ }\href@noop {} {\bibfield  {journal}
  {\bibinfo  {journal} {Neural networks}\ }\textbf {\bibinfo {volume} {2}},\
  \bibinfo {pages} {359} (\bibinfo {year} {1989})}\BibitemShut {NoStop}%
\bibitem [{\citenamefont {Chen}\ \emph {et~al.}(2018)\citenamefont {Chen},
  \citenamefont {Rubanova}, \citenamefont {Bettencourt},\ and\ \citenamefont
  {Duvenaud}}]{chen2018neuralode}%
  \BibitemOpen
  \bibfield  {author} {\bibinfo {author} {\bibfnamefont {R.~T.~Q.}\
  \bibnamefont {Chen}}, \bibinfo {author} {\bibfnamefont {Y.}~\bibnamefont
  {Rubanova}}, \bibinfo {author} {\bibfnamefont {J.}~\bibnamefont
  {Bettencourt}},\ and\ \bibinfo {author} {\bibfnamefont {D.}~\bibnamefont
  {Duvenaud}},\ }\bibfield  {title} {\bibinfo {title} {Neural ordinary
  differential equations},\ }\href@noop {} {\bibfield  {journal} {\bibinfo
  {journal} {Advances in Neural Information Processing Systems}\ } (\bibinfo
  {year} {2018})}\BibitemShut {NoStop}%
\bibitem [{\citenamefont {Sauvage}\ and\ \citenamefont
  {Mintert}(2022)}]{sauvage2022optimal}%
  \BibitemOpen
  \bibfield  {author} {\bibinfo {author} {\bibfnamefont {F.}~\bibnamefont
  {Sauvage}}\ and\ \bibinfo {author} {\bibfnamefont {F.}~\bibnamefont
  {Mintert}},\ }\bibfield  {title} {\bibinfo {title} {Optimal control of
  families of quantum gates},\ }\href@noop {} {\bibfield  {journal} {\bibinfo
  {journal} {Physical review letters}\ }\textbf {\bibinfo {volume} {129}},\
  \bibinfo {pages} {050507} (\bibinfo {year} {2022})}\BibitemShut {NoStop}%
\bibitem [{\citenamefont {G{\"u}ng{\"o}rd{\"u}}\ and\ \citenamefont
  {Kestner}(2022)}]{gungordu2022robust}%
  \BibitemOpen
  \bibfield  {author} {\bibinfo {author} {\bibfnamefont {U.}~\bibnamefont
  {G{\"u}ng{\"o}rd{\"u}}}\ and\ \bibinfo {author} {\bibfnamefont
  {J.}~\bibnamefont {Kestner}},\ }\bibfield  {title} {\bibinfo {title} {Robust
  quantum gates using smooth pulses and physics-informed neural networks},\
  }\href@noop {} {\bibfield  {journal} {\bibinfo  {journal} {Physical Review
  Research}\ }\textbf {\bibinfo {volume} {4}},\ \bibinfo {pages} {023155}
  (\bibinfo {year} {2022})}\BibitemShut {NoStop}%
\bibitem [{\citenamefont {Bhattacharyya}\ \emph {et~al.}(2024)\citenamefont
  {Bhattacharyya}, \citenamefont {An}, \citenamefont {Kozbiel}, \citenamefont
  {Goldschmidt},\ and\ \citenamefont {Chong}}]{bhattacharyya2024using}%
  \BibitemOpen
  \bibfield  {author} {\bibinfo {author} {\bibfnamefont {B.}~\bibnamefont
  {Bhattacharyya}}, \bibinfo {author} {\bibfnamefont {F.}~\bibnamefont {An}},
  \bibinfo {author} {\bibfnamefont {D.}~\bibnamefont {Kozbiel}}, \bibinfo
  {author} {\bibfnamefont {A.~J.}\ \bibnamefont {Goldschmidt}},\ and\ \bibinfo
  {author} {\bibfnamefont {F.~T.}\ \bibnamefont {Chong}},\ }\bibfield  {title}
  {\bibinfo {title} {Using optimal control to guide neural-network
  interpolation of continuously-parameterized gates},\ }in\ \href@noop {}
  {\emph {\bibinfo {booktitle} {2024 IEEE International Conference on Quantum
  Computing and Engineering (QCE)}}},\ Vol.~\bibinfo {volume} {1}\ (\bibinfo
  {organization} {IEEE},\ \bibinfo {year} {2024})\ pp.\ \bibinfo {pages}
  {1336--1346}\BibitemShut {NoStop}%
\bibitem [{\citenamefont {McCord}\ \emph {et~al.}(2024)\citenamefont {McCord},
  \citenamefont {Dogra},\ and\ \citenamefont {Paraoanu}}]{McCord2024}%
  \BibitemOpen
  \bibfield  {author} {\bibinfo {author} {\bibfnamefont {J.~J.}\ \bibnamefont
  {McCord}}, \bibinfo {author} {\bibfnamefont {S.}~\bibnamefont {Dogra}},\ and\
  \bibinfo {author} {\bibfnamefont {G.~S.}\ \bibnamefont {Paraoanu}},\
  }\bibfield  {title} {\bibinfo {title} {Coherent interaction-free detection of
  noise},\ }\href {https://doi.org/10.1103/PhysRevA.110.032404} {\bibfield
  {journal} {\bibinfo  {journal} {Phys. Rev. A}\ }\textbf {\bibinfo {volume}
  {110}},\ \bibinfo {pages} {032404} (\bibinfo {year} {2024})}\BibitemShut
  {NoStop}%
\bibitem [{\citenamefont {McCord}\ \emph {et~al.}(2023)\citenamefont {McCord},
  \citenamefont {Dogra},\ and\ \citenamefont {Paraoanu}}]{McCord2023}%
  \BibitemOpen
  \bibfield  {author} {\bibinfo {author} {\bibfnamefont {J.~J.}\ \bibnamefont
  {McCord}}, \bibinfo {author} {\bibfnamefont {S.}~\bibnamefont {Dogra}},\ and\
  \bibinfo {author} {\bibfnamefont {G.~S.}\ \bibnamefont {Paraoanu}},\
  }\bibfield  {title} {\bibinfo {title} {Theory of coherent interaction-free
  detection of pulses},\ }\href
  {https://doi.org/10.1103/PhysRevResearch.5.033012} {\bibfield  {journal}
  {\bibinfo  {journal} {Phys. Rev. Res.}\ }\textbf {\bibinfo {volume} {5}},\
  \bibinfo {pages} {033012} (\bibinfo {year} {2023})}\BibitemShut {NoStop}%
\bibitem [{\citenamefont {Dogra}\ \emph {et~al.}(2022)\citenamefont {Dogra},
  \citenamefont {McCord},\ and\ \citenamefont {Paraoanu}}]{Dogra2022}%
  \BibitemOpen
  \bibfield  {author} {\bibinfo {author} {\bibfnamefont {S.}~\bibnamefont
  {Dogra}}, \bibinfo {author} {\bibfnamefont {J.~J.}\ \bibnamefont {McCord}},\
  and\ \bibinfo {author} {\bibfnamefont {G.~S.}\ \bibnamefont {Paraoanu}},\
  }\bibfield  {title} {\bibinfo {title} {Coherent interaction-free detection of
  microwave pulses with a superconducting circuit},\ }\href
  {https://doi.org/10.1038/s41467-022-35049-z} {\bibfield  {journal} {\bibinfo
  {journal} {Nat. Commun.}\ }\textbf {\bibinfo {volume} {13}},\ \bibinfo
  {pages} {7528} (\bibinfo {year} {2022})}\BibitemShut {NoStop}%
\bibitem [{\citenamefont {Dixit}\ \emph {et~al.}(2021)\citenamefont {Dixit},
  \citenamefont {Chakram}, \citenamefont {He}, \citenamefont {Agrawal},
  \citenamefont {Naik}, \citenamefont {Schuster},\ and\ \citenamefont
  {Chou}}]{Dixit2021}%
  \BibitemOpen
  \bibfield  {author} {\bibinfo {author} {\bibfnamefont {A.~V.}\ \bibnamefont
  {Dixit}}, \bibinfo {author} {\bibfnamefont {S.}~\bibnamefont {Chakram}},
  \bibinfo {author} {\bibfnamefont {K.}~\bibnamefont {He}}, \bibinfo {author}
  {\bibfnamefont {A.}~\bibnamefont {Agrawal}}, \bibinfo {author} {\bibfnamefont
  {R.~K.}\ \bibnamefont {Naik}}, \bibinfo {author} {\bibfnamefont {D.~I.}\
  \bibnamefont {Schuster}},\ and\ \bibinfo {author} {\bibfnamefont
  {A.}~\bibnamefont {Chou}},\ }\bibfield  {title} {\bibinfo {title} {Searching
  for dark matter with a superconducting qubit},\ }\href
  {https://doi.org/10.1103/PhysRevLett.126.141302} {\bibfield  {journal}
  {\bibinfo  {journal} {Phys. Rev. Lett.}\ }\textbf {\bibinfo {volume} {126}},\
  \bibinfo {pages} {141302} (\bibinfo {year} {2021})}\BibitemShut {NoStop}%
\bibitem [{\citenamefont {Gusarov}\ \emph {et~al.}(2023)\citenamefont
  {Gusarov}, \citenamefont {Perelshtein}, \citenamefont {Hakonen},\ and\
  \citenamefont {Paraoanu}}]{Gusarov2023}%
  \BibitemOpen
  \bibfield  {author} {\bibinfo {author} {\bibfnamefont {N.~N.}\ \bibnamefont
  {Gusarov}}, \bibinfo {author} {\bibfnamefont {M.~R.}\ \bibnamefont
  {Perelshtein}}, \bibinfo {author} {\bibfnamefont {P.~J.}\ \bibnamefont
  {Hakonen}},\ and\ \bibinfo {author} {\bibfnamefont {G.~S.}\ \bibnamefont
  {Paraoanu}},\ }\bibfield  {title} {\bibinfo {title} {Optimized emulation of
  quantum magnetometry via superconducting qubits},\ }\href
  {https://doi.org/10.1103/PhysRevA.107.052609} {\bibfield  {journal} {\bibinfo
   {journal} {Phys. Rev. A}\ }\textbf {\bibinfo {volume} {107}},\ \bibinfo
  {pages} {052609} (\bibinfo {year} {2023})}\BibitemShut {NoStop}%
\bibitem [{\citenamefont {Shlyakhov}\ \emph {et~al.}(2018)\citenamefont
  {Shlyakhov}, \citenamefont {Zemlyanov}, \citenamefont {Suslov}, \citenamefont
  {Lebedev}, \citenamefont {Paraoanu}, \citenamefont {Lesovik},\ and\
  \citenamefont {Blatter}}]{Blatter2018}%
  \BibitemOpen
  \bibfield  {author} {\bibinfo {author} {\bibfnamefont {A.~R.}\ \bibnamefont
  {Shlyakhov}}, \bibinfo {author} {\bibfnamefont {V.~V.}\ \bibnamefont
  {Zemlyanov}}, \bibinfo {author} {\bibfnamefont {M.~V.}\ \bibnamefont
  {Suslov}}, \bibinfo {author} {\bibfnamefont {A.~V.}\ \bibnamefont {Lebedev}},
  \bibinfo {author} {\bibfnamefont {G.~S.}\ \bibnamefont {Paraoanu}}, \bibinfo
  {author} {\bibfnamefont {G.~B.}\ \bibnamefont {Lesovik}},\ and\ \bibinfo
  {author} {\bibfnamefont {G.}~\bibnamefont {Blatter}},\ }\bibfield  {title}
  {\bibinfo {title} {Quantum metrology with a transmon qutrit},\ }\href
  {https://doi.org/10.1103/PhysRevA.97.022115} {\bibfield  {journal} {\bibinfo
  {journal} {Phys. Rev. A}\ }\textbf {\bibinfo {volume} {97}},\ \bibinfo
  {pages} {022115} (\bibinfo {year} {2018})}\BibitemShut {NoStop}%
\bibitem [{\citenamefont {Danilin}\ \emph {et~al.}(2018)\citenamefont
  {Danilin}, \citenamefont {Lebedev}, \citenamefont {Vepsäläinen},
  \citenamefont {Lesovik}, \citenamefont {Blatter},\ and\ \citenamefont
  {Paraoanu}}]{Danilin2018}%
  \BibitemOpen
  \bibfield  {author} {\bibinfo {author} {\bibfnamefont {S.}~\bibnamefont
  {Danilin}}, \bibinfo {author} {\bibfnamefont {A.~V.}\ \bibnamefont
  {Lebedev}}, \bibinfo {author} {\bibfnamefont {A.}~\bibnamefont
  {Vepsäläinen}}, \bibinfo {author} {\bibfnamefont {G.~B.}\ \bibnamefont
  {Lesovik}}, \bibinfo {author} {\bibfnamefont {G.}~\bibnamefont {Blatter}},\
  and\ \bibinfo {author} {\bibfnamefont {G.~S.}\ \bibnamefont {Paraoanu}},\
  }\bibfield  {title} {\bibinfo {title} {Quantum-enhanced magnetometry by phase
  estimation algorithms with a single artificial atom},\ }\href
  {https://doi.org/10.1038/s41534-018-0078-y} {\bibfield  {journal} {\bibinfo
  {journal} {npj Quantum Information}\ }\textbf {\bibinfo {volume} {4}},\
  \bibinfo {pages} {29} (\bibinfo {year} {2018})}\BibitemShut {NoStop}%
\bibitem [{\citenamefont {Milul}\ \emph {et~al.}(2023)\citenamefont {Milul},
  \citenamefont {Guttel}, \citenamefont {Goldblatt}, \citenamefont {Hazanov},
  \citenamefont {Joshi}, \citenamefont {Chausovsky}, \citenamefont {Kahn},
  \citenamefont {\ifmmode~\mbox{\c{C}}\else \c{C}\fi{}ifty\"urek},
  \citenamefont {Lafont},\ and\ \citenamefont
  {Rosenblum}}]{PRXQuantum.4.030336}%
  \BibitemOpen
  \bibfield  {author} {\bibinfo {author} {\bibfnamefont {O.}~\bibnamefont
  {Milul}}, \bibinfo {author} {\bibfnamefont {B.}~\bibnamefont {Guttel}},
  \bibinfo {author} {\bibfnamefont {U.}~\bibnamefont {Goldblatt}}, \bibinfo
  {author} {\bibfnamefont {S.}~\bibnamefont {Hazanov}}, \bibinfo {author}
  {\bibfnamefont {L.~M.}\ \bibnamefont {Joshi}}, \bibinfo {author}
  {\bibfnamefont {D.}~\bibnamefont {Chausovsky}}, \bibinfo {author}
  {\bibfnamefont {N.}~\bibnamefont {Kahn}}, \bibinfo {author} {\bibfnamefont
  {E.}~\bibnamefont {\ifmmode~\mbox{\c{C}}\else \c{C}\fi{}ifty\"urek}},
  \bibinfo {author} {\bibfnamefont {F.}~\bibnamefont {Lafont}},\ and\ \bibinfo
  {author} {\bibfnamefont {S.}~\bibnamefont {Rosenblum}},\ }\bibfield  {title}
  {\bibinfo {title} {Superconducting cavity qubit with tens of milliseconds
  single-photon coherence time},\ }\href
  {https://doi.org/10.1103/PRXQuantum.4.030336} {\bibfield  {journal} {\bibinfo
   {journal} {PRX Quantum}\ }\textbf {\bibinfo {volume} {4}},\ \bibinfo {pages}
  {030336} (\bibinfo {year} {2023})}\BibitemShut {NoStop}%
\bibitem [{\citenamefont {Bertet}\ \emph {et~al.}(2002)\citenamefont {Bertet},
  \citenamefont {Auffeves}, \citenamefont {Maioli}, \citenamefont {Osnaghi},
  \citenamefont {Meunier}, \citenamefont {Brune}, \citenamefont {Raimond},\
  and\ \citenamefont {Haroche}}]{PhysRevLett.89.200402}%
  \BibitemOpen
  \bibfield  {author} {\bibinfo {author} {\bibfnamefont {P.}~\bibnamefont
  {Bertet}}, \bibinfo {author} {\bibfnamefont {A.}~\bibnamefont {Auffeves}},
  \bibinfo {author} {\bibfnamefont {P.}~\bibnamefont {Maioli}}, \bibinfo
  {author} {\bibfnamefont {S.}~\bibnamefont {Osnaghi}}, \bibinfo {author}
  {\bibfnamefont {T.}~\bibnamefont {Meunier}}, \bibinfo {author} {\bibfnamefont
  {M.}~\bibnamefont {Brune}}, \bibinfo {author} {\bibfnamefont {J.~M.}\
  \bibnamefont {Raimond}},\ and\ \bibinfo {author} {\bibfnamefont
  {S.}~\bibnamefont {Haroche}},\ }\bibfield  {title} {\bibinfo {title} {Direct
  measurement of the wigner function of a one-photon fock state in a cavity},\
  }\href {https://doi.org/10.1103/PhysRevLett.89.200402} {\bibfield  {journal}
  {\bibinfo  {journal} {Phys. Rev. Lett.}\ }\textbf {\bibinfo {volume} {89}},\
  \bibinfo {pages} {200402} (\bibinfo {year} {2002})}\BibitemShut {NoStop}%
\bibitem [{\citenamefont {Vlastakis}\ \emph {et~al.}(2013)\citenamefont
  {Vlastakis}, \citenamefont {Kirchmair}, \citenamefont {Leghtas},
  \citenamefont {Nigg}, \citenamefont {Frunzio}, \citenamefont {Girvin},
  \citenamefont {Mirrahimi}, \citenamefont {Devoret},\ and\ \citenamefont
  {Schoelkopf}}]{doi:10.1126/science.1243289}%
  \BibitemOpen
  \bibfield  {author} {\bibinfo {author} {\bibfnamefont {B.}~\bibnamefont
  {Vlastakis}}, \bibinfo {author} {\bibfnamefont {G.}~\bibnamefont
  {Kirchmair}}, \bibinfo {author} {\bibfnamefont {Z.}~\bibnamefont {Leghtas}},
  \bibinfo {author} {\bibfnamefont {S.~E.}\ \bibnamefont {Nigg}}, \bibinfo
  {author} {\bibfnamefont {L.}~\bibnamefont {Frunzio}}, \bibinfo {author}
  {\bibfnamefont {S.~M.}\ \bibnamefont {Girvin}}, \bibinfo {author}
  {\bibfnamefont {M.}~\bibnamefont {Mirrahimi}}, \bibinfo {author}
  {\bibfnamefont {M.~H.}\ \bibnamefont {Devoret}},\ and\ \bibinfo {author}
  {\bibfnamefont {R.~J.}\ \bibnamefont {Schoelkopf}},\ }\bibfield  {title}
  {\bibinfo {title} {Deterministically encoding quantum information using
  100-photon schrödinger cat states},\ }\href
  {https://doi.org/10.1126/science.1243289} {\bibfield  {journal} {\bibinfo
  {journal} {Science}\ }\textbf {\bibinfo {volume} {342}},\ \bibinfo {pages}
  {607} (\bibinfo {year} {2013})},\ \Eprint
  {https://arxiv.org/abs/https://www.science.org/doi/pdf/10.1126/science.1243289}
  {https://www.science.org/doi/pdf/10.1126/science.1243289} \BibitemShut
  {NoStop}%
\bibitem [{\citenamefont {Sun}\ \emph {et~al.}(2014)\citenamefont {Sun},
  \citenamefont {Petrenko}, \citenamefont {Leghtas}, \citenamefont {Vlastakis},
  \citenamefont {Kirchmair}, \citenamefont {Sliwa}, \citenamefont {Narla},
  \citenamefont {Hatridge}, \citenamefont {Shankar}, \citenamefont {Blumoff},
  \citenamefont {Frunzio}, \citenamefont {Mirrahimi}, \citenamefont {Devoret},\
  and\ \citenamefont {Schoelkopf}}]{Sun2014}%
  \BibitemOpen
  \bibfield  {author} {\bibinfo {author} {\bibfnamefont {L.}~\bibnamefont
  {Sun}}, \bibinfo {author} {\bibfnamefont {A.}~\bibnamefont {Petrenko}},
  \bibinfo {author} {\bibfnamefont {Z.}~\bibnamefont {Leghtas}}, \bibinfo
  {author} {\bibfnamefont {B.}~\bibnamefont {Vlastakis}}, \bibinfo {author}
  {\bibfnamefont {G.}~\bibnamefont {Kirchmair}}, \bibinfo {author}
  {\bibfnamefont {K.~M.}\ \bibnamefont {Sliwa}}, \bibinfo {author}
  {\bibfnamefont {A.}~\bibnamefont {Narla}}, \bibinfo {author} {\bibfnamefont
  {M.}~\bibnamefont {Hatridge}}, \bibinfo {author} {\bibfnamefont
  {S.}~\bibnamefont {Shankar}}, \bibinfo {author} {\bibfnamefont
  {J.}~\bibnamefont {Blumoff}}, \bibinfo {author} {\bibfnamefont
  {L.}~\bibnamefont {Frunzio}}, \bibinfo {author} {\bibfnamefont
  {M.}~\bibnamefont {Mirrahimi}}, \bibinfo {author} {\bibfnamefont {M.~H.}\
  \bibnamefont {Devoret}},\ and\ \bibinfo {author} {\bibfnamefont {R.~J.}\
  \bibnamefont {Schoelkopf}},\ }\bibfield  {title} {\bibinfo {title} {Tracking
  photon jumps with repeated quantum non-demolition parity measurements},\
  }\href {https://doi.org/10.1038/nature13436} {\bibfield  {journal} {\bibinfo
  {journal} {Nature}\ }\textbf {\bibinfo {volume} {511}},\ \bibinfo {pages}
  {444} (\bibinfo {year} {2014})}\BibitemShut {NoStop}%
\bibitem [{\citenamefont {Chen}(2018)}]{torchdiffeq}%
  \BibitemOpen
  \bibfield  {author} {\bibinfo {author} {\bibfnamefont {R.~T.~Q.}\
  \bibnamefont {Chen}},\ }\href {https://github.com/rtqichen/torchdiffeq}
  {\bibinfo {title} {torchdiffeq}} (\bibinfo {year} {2018})\BibitemShut
  {NoStop}%
\bibitem [{\citenamefont {Diederik}(2014)}]{diederik2014adam}%
  \BibitemOpen
  \bibfield  {author} {\bibinfo {author} {\bibfnamefont {K.}~\bibnamefont
  {Diederik}},\ }\bibfield  {title} {\bibinfo {title} {Adam: A method for
  stochastic optimization},\ }\href@noop {} {\bibfield  {journal} {\bibinfo
  {journal} {(No Title)}\ } (\bibinfo {year} {2014})}\BibitemShut {NoStop}%
\bibitem [{\citenamefont {Osman}\ \emph {et~al.}(2023)\citenamefont {Osman},
  \citenamefont {Fern\'andez-Pend\'as}, \citenamefont {Warren}, \citenamefont
  {Kosen}, \citenamefont {Scigliuzzo}, \citenamefont {Frisk~Kockum},
  \citenamefont {Tancredi}, \citenamefont {Fadavi~Roudsari},\ and\
  \citenamefont {Bylander}}]{PhysRevResearch.5.043001}%
  \BibitemOpen
  \bibfield  {author} {\bibinfo {author} {\bibfnamefont {A.}~\bibnamefont
  {Osman}}, \bibinfo {author} {\bibfnamefont {J.}~\bibnamefont
  {Fern\'andez-Pend\'as}}, \bibinfo {author} {\bibfnamefont {C.}~\bibnamefont
  {Warren}}, \bibinfo {author} {\bibfnamefont {S.}~\bibnamefont {Kosen}},
  \bibinfo {author} {\bibfnamefont {M.}~\bibnamefont {Scigliuzzo}}, \bibinfo
  {author} {\bibfnamefont {A.}~\bibnamefont {Frisk~Kockum}}, \bibinfo {author}
  {\bibfnamefont {G.}~\bibnamefont {Tancredi}}, \bibinfo {author}
  {\bibfnamefont {A.}~\bibnamefont {Fadavi~Roudsari}},\ and\ \bibinfo {author}
  {\bibfnamefont {J.}~\bibnamefont {Bylander}},\ }\bibfield  {title} {\bibinfo
  {title} {Mitigation of frequency collisions in superconducting quantum
  processors},\ }\href {https://doi.org/10.1103/PhysRevResearch.5.043001}
  {\bibfield  {journal} {\bibinfo  {journal} {Phys. Rev. Res.}\ }\textbf
  {\bibinfo {volume} {5}},\ \bibinfo {pages} {043001} (\bibinfo {year}
  {2023})}\BibitemShut {NoStop}%
\bibitem [{\citenamefont {Magesan}\ \emph {et~al.}(2012)\citenamefont
  {Magesan}, \citenamefont {Gambetta}, \citenamefont {Johnson}, \citenamefont
  {Ryan}, \citenamefont {Chow}, \citenamefont {Merkel}, \citenamefont
  {da~Silva}, \citenamefont {Keefe}, \citenamefont {Rothwell}, \citenamefont
  {Ohki}, \citenamefont {Ketchen},\ and\ \citenamefont
  {Steffen}}]{PhysRevLett.109.080505}%
  \BibitemOpen
  \bibfield  {author} {\bibinfo {author} {\bibfnamefont {E.}~\bibnamefont
  {Magesan}}, \bibinfo {author} {\bibfnamefont {J.~M.}\ \bibnamefont
  {Gambetta}}, \bibinfo {author} {\bibfnamefont {B.~R.}\ \bibnamefont
  {Johnson}}, \bibinfo {author} {\bibfnamefont {C.~A.}\ \bibnamefont {Ryan}},
  \bibinfo {author} {\bibfnamefont {J.~M.}\ \bibnamefont {Chow}}, \bibinfo
  {author} {\bibfnamefont {S.~T.}\ \bibnamefont {Merkel}}, \bibinfo {author}
  {\bibfnamefont {M.~P.}\ \bibnamefont {da~Silva}}, \bibinfo {author}
  {\bibfnamefont {G.~A.}\ \bibnamefont {Keefe}}, \bibinfo {author}
  {\bibfnamefont {M.~B.}\ \bibnamefont {Rothwell}}, \bibinfo {author}
  {\bibfnamefont {T.~A.}\ \bibnamefont {Ohki}}, \bibinfo {author}
  {\bibfnamefont {M.~B.}\ \bibnamefont {Ketchen}},\ and\ \bibinfo {author}
  {\bibfnamefont {M.}~\bibnamefont {Steffen}},\ }\bibfield  {title} {\bibinfo
  {title} {Efficient measurement of quantum gate error by interleaved
  randomized benchmarking},\ }\href
  {https://doi.org/10.1103/PhysRevLett.109.080505} {\bibfield  {journal}
  {\bibinfo  {journal} {Phys. Rev. Lett.}\ }\textbf {\bibinfo {volume} {109}},\
  \bibinfo {pages} {080505} (\bibinfo {year} {2012})}\BibitemShut {NoStop}%
\bibitem [{\citenamefont {Magesan}\ \emph {et~al.}(2011)\citenamefont
  {Magesan}, \citenamefont {Gambetta},\ and\ \citenamefont
  {Emerson}}]{PhysRevLett.106.180504}%
  \BibitemOpen
  \bibfield  {author} {\bibinfo {author} {\bibfnamefont {E.}~\bibnamefont
  {Magesan}}, \bibinfo {author} {\bibfnamefont {J.~M.}\ \bibnamefont
  {Gambetta}},\ and\ \bibinfo {author} {\bibfnamefont {J.}~\bibnamefont
  {Emerson}},\ }\bibfield  {title} {\bibinfo {title} {Scalable and robust
  randomized benchmarking of quantum processes},\ }\href
  {https://doi.org/10.1103/PhysRevLett.106.180504} {\bibfield  {journal}
  {\bibinfo  {journal} {Phys. Rev. Lett.}\ }\textbf {\bibinfo {volume} {106}},\
  \bibinfo {pages} {180504} (\bibinfo {year} {2011})}\BibitemShut {NoStop}%
\bibitem [{\citenamefont {Bronzan}(1988)}]{bronzan1988parametrization}%
  \BibitemOpen
  \bibfield  {author} {\bibinfo {author} {\bibfnamefont {J.~B.}\ \bibnamefont
  {Bronzan}},\ }\bibfield  {title} {\bibinfo {title} {Parametrization of su
  (3)},\ }\href@noop {} {\bibfield  {journal} {\bibinfo  {journal} {Physical
  Review D}\ }\textbf {\bibinfo {volume} {38}},\ \bibinfo {pages} {1994}
  (\bibinfo {year} {1988})}\BibitemShut {NoStop}%
\bibitem [{\citenamefont {Ding}\ \emph {et~al.}(2023)\citenamefont {Ding},
  \citenamefont {Hays}, \citenamefont {Sung}, \citenamefont {Kannan},
  \citenamefont {An}, \citenamefont {Di~Paolo}, \citenamefont {Karamlou},
  \citenamefont {Hazard}, \citenamefont {Azar}, \citenamefont {Kim},
  \citenamefont {Niedzielski}, \citenamefont {Melville}, \citenamefont
  {Schwartz}, \citenamefont {Yoder}, \citenamefont {Orlando}, \citenamefont
  {Gustavsson}, \citenamefont {Grover}, \citenamefont {Serniak},\ and\
  \citenamefont {Oliver}}]{PhysRevX.13.031035}%
  \BibitemOpen
  \bibfield  {author} {\bibinfo {author} {\bibfnamefont {L.}~\bibnamefont
  {Ding}}, \bibinfo {author} {\bibfnamefont {M.}~\bibnamefont {Hays}}, \bibinfo
  {author} {\bibfnamefont {Y.}~\bibnamefont {Sung}}, \bibinfo {author}
  {\bibfnamefont {B.}~\bibnamefont {Kannan}}, \bibinfo {author} {\bibfnamefont
  {J.}~\bibnamefont {An}}, \bibinfo {author} {\bibfnamefont {A.}~\bibnamefont
  {Di~Paolo}}, \bibinfo {author} {\bibfnamefont {A.~H.}\ \bibnamefont
  {Karamlou}}, \bibinfo {author} {\bibfnamefont {T.~M.}\ \bibnamefont
  {Hazard}}, \bibinfo {author} {\bibfnamefont {K.}~\bibnamefont {Azar}},
  \bibinfo {author} {\bibfnamefont {D.~K.}\ \bibnamefont {Kim}}, \bibinfo
  {author} {\bibfnamefont {B.~M.}\ \bibnamefont {Niedzielski}}, \bibinfo
  {author} {\bibfnamefont {A.}~\bibnamefont {Melville}}, \bibinfo {author}
  {\bibfnamefont {M.~E.}\ \bibnamefont {Schwartz}}, \bibinfo {author}
  {\bibfnamefont {J.~L.}\ \bibnamefont {Yoder}}, \bibinfo {author}
  {\bibfnamefont {T.~P.}\ \bibnamefont {Orlando}}, \bibinfo {author}
  {\bibfnamefont {S.}~\bibnamefont {Gustavsson}}, \bibinfo {author}
  {\bibfnamefont {J.~A.}\ \bibnamefont {Grover}}, \bibinfo {author}
  {\bibfnamefont {K.}~\bibnamefont {Serniak}},\ and\ \bibinfo {author}
  {\bibfnamefont {W.~D.}\ \bibnamefont {Oliver}},\ }\bibfield  {title}
  {\bibinfo {title} {High-fidelity, frequency-flexible two-qubit fluxonium
  gates with a transmon coupler},\ }\href
  {https://doi.org/10.1103/PhysRevX.13.031035} {\bibfield  {journal} {\bibinfo
  {journal} {Phys. Rev. X}\ }\textbf {\bibinfo {volume} {13}},\ \bibinfo
  {pages} {031035} (\bibinfo {year} {2023})}\BibitemShut {NoStop}%
\bibitem [{\citenamefont {Abad}\ \emph {et~al.}(2022)\citenamefont {Abad},
  \citenamefont {Fern{\'a}ndez-Pend{\'a}s}, \citenamefont {Frisk~Kockum},\ and\
  \citenamefont {Johansson}}]{abad2022universal}%
  \BibitemOpen
  \bibfield  {author} {\bibinfo {author} {\bibfnamefont {T.}~\bibnamefont
  {Abad}}, \bibinfo {author} {\bibfnamefont {J.}~\bibnamefont
  {Fern{\'a}ndez-Pend{\'a}s}}, \bibinfo {author} {\bibfnamefont
  {A.}~\bibnamefont {Frisk~Kockum}},\ and\ \bibinfo {author} {\bibfnamefont
  {G.}~\bibnamefont {Johansson}},\ }\bibfield  {title} {\bibinfo {title}
  {Universal fidelity reduction of quantum operations from weak dissipation},\
  }\href@noop {} {\bibfield  {journal} {\bibinfo  {journal} {Physical Review
  Letters}\ }\textbf {\bibinfo {volume} {129}},\ \bibinfo {pages} {150504}
  (\bibinfo {year} {2022})}\BibitemShut {NoStop}%
\bibitem [{\citenamefont {Bengtsson}\ \emph {et~al.}(2020)\citenamefont
  {Bengtsson}, \citenamefont {Vikst\aa{}l}, \citenamefont {Warren},
  \citenamefont {Svensson}, \citenamefont {Gu}, \citenamefont {Kockum},
  \citenamefont {Krantz}, \citenamefont {Kri\ifmmode~\check{z}\else
  \v{z}\fi{}an}, \citenamefont {Shiri}, \citenamefont {Svensson}, \citenamefont
  {Tancredi}, \citenamefont {Johansson}, \citenamefont {Delsing}, \citenamefont
  {Ferrini},\ and\ \citenamefont {Bylander}}]{PhysRevApplied.14.034010}%
  \BibitemOpen
  \bibfield  {author} {\bibinfo {author} {\bibfnamefont {A.}~\bibnamefont
  {Bengtsson}}, \bibinfo {author} {\bibfnamefont {P.}~\bibnamefont
  {Vikst\aa{}l}}, \bibinfo {author} {\bibfnamefont {C.}~\bibnamefont {Warren}},
  \bibinfo {author} {\bibfnamefont {M.}~\bibnamefont {Svensson}}, \bibinfo
  {author} {\bibfnamefont {X.}~\bibnamefont {Gu}}, \bibinfo {author}
  {\bibfnamefont {A.~F.}\ \bibnamefont {Kockum}}, \bibinfo {author}
  {\bibfnamefont {P.}~\bibnamefont {Krantz}}, \bibinfo {author} {\bibfnamefont
  {C.}~\bibnamefont {Kri\ifmmode~\check{z}\else \v{z}\fi{}an}}, \bibinfo
  {author} {\bibfnamefont {D.}~\bibnamefont {Shiri}}, \bibinfo {author}
  {\bibfnamefont {I.-M.}\ \bibnamefont {Svensson}}, \bibinfo {author}
  {\bibfnamefont {G.}~\bibnamefont {Tancredi}}, \bibinfo {author}
  {\bibfnamefont {G.}~\bibnamefont {Johansson}}, \bibinfo {author}
  {\bibfnamefont {P.}~\bibnamefont {Delsing}}, \bibinfo {author} {\bibfnamefont
  {G.}~\bibnamefont {Ferrini}},\ and\ \bibinfo {author} {\bibfnamefont
  {J.}~\bibnamefont {Bylander}},\ }\bibfield  {title} {\bibinfo {title}
  {Improved success probability with greater circuit depth for the quantum
  approximate optimization algorithm},\ }\href
  {https://doi.org/10.1103/PhysRevApplied.14.034010} {\bibfield  {journal}
  {\bibinfo  {journal} {Phys. Rev. Appl.}\ }\textbf {\bibinfo {volume} {14}},\
  \bibinfo {pages} {034010} (\bibinfo {year} {2020})}\BibitemShut {NoStop}%
\bibitem [{\citenamefont {Chen}\ \emph {et~al.}(2016)\citenamefont {Chen},
  \citenamefont {Kelly}, \citenamefont {Quintana}, \citenamefont {Barends},
  \citenamefont {Campbell}, \citenamefont {Chen}, \citenamefont {Chiaro},
  \citenamefont {Dunsworth}, \citenamefont {Fowler}, \citenamefont {Lucero},
  \citenamefont {Jeffrey}, \citenamefont {Megrant}, \citenamefont {Mutus},
  \citenamefont {Neeley}, \citenamefont {Neill}, \citenamefont {O'Malley},
  \citenamefont {Roushan}, \citenamefont {Sank}, \citenamefont {Vainsencher},
  \citenamefont {Wenner}, \citenamefont {White}, \citenamefont {Korotkov},\
  and\ \citenamefont {Martinis}}]{PhysRevLett.116.020501}%
  \BibitemOpen
  \bibfield  {author} {\bibinfo {author} {\bibfnamefont {Z.}~\bibnamefont
  {Chen}}, \bibinfo {author} {\bibfnamefont {J.}~\bibnamefont {Kelly}},
  \bibinfo {author} {\bibfnamefont {C.}~\bibnamefont {Quintana}}, \bibinfo
  {author} {\bibfnamefont {R.}~\bibnamefont {Barends}}, \bibinfo {author}
  {\bibfnamefont {B.}~\bibnamefont {Campbell}}, \bibinfo {author}
  {\bibfnamefont {Y.}~\bibnamefont {Chen}}, \bibinfo {author} {\bibfnamefont
  {B.}~\bibnamefont {Chiaro}}, \bibinfo {author} {\bibfnamefont
  {A.}~\bibnamefont {Dunsworth}}, \bibinfo {author} {\bibfnamefont {A.~G.}\
  \bibnamefont {Fowler}}, \bibinfo {author} {\bibfnamefont {E.}~\bibnamefont
  {Lucero}}, \bibinfo {author} {\bibfnamefont {E.}~\bibnamefont {Jeffrey}},
  \bibinfo {author} {\bibfnamefont {A.}~\bibnamefont {Megrant}}, \bibinfo
  {author} {\bibfnamefont {J.}~\bibnamefont {Mutus}}, \bibinfo {author}
  {\bibfnamefont {M.}~\bibnamefont {Neeley}}, \bibinfo {author} {\bibfnamefont
  {C.}~\bibnamefont {Neill}}, \bibinfo {author} {\bibfnamefont {P.~J.~J.}\
  \bibnamefont {O'Malley}}, \bibinfo {author} {\bibfnamefont {P.}~\bibnamefont
  {Roushan}}, \bibinfo {author} {\bibfnamefont {D.}~\bibnamefont {Sank}},
  \bibinfo {author} {\bibfnamefont {A.}~\bibnamefont {Vainsencher}}, \bibinfo
  {author} {\bibfnamefont {J.}~\bibnamefont {Wenner}}, \bibinfo {author}
  {\bibfnamefont {T.~C.}\ \bibnamefont {White}}, \bibinfo {author}
  {\bibfnamefont {A.~N.}\ \bibnamefont {Korotkov}},\ and\ \bibinfo {author}
  {\bibfnamefont {J.~M.}\ \bibnamefont {Martinis}},\ }\bibfield  {title}
  {\bibinfo {title} {Measuring and suppressing quantum state leakage in a
  superconducting qubit},\ }\href
  {https://doi.org/10.1103/PhysRevLett.116.020501} {\bibfield  {journal}
  {\bibinfo  {journal} {Phys. Rev. Lett.}\ }\textbf {\bibinfo {volume} {116}},\
  \bibinfo {pages} {020501} (\bibinfo {year} {2016})}\BibitemShut {NoStop}%
\bibitem [{\citenamefont {Schreier}\ and\ \citenamefont
  {Scharf}(2010)}]{schreier2010statistical}%
  \BibitemOpen
  \bibfield  {author} {\bibinfo {author} {\bibfnamefont {P.~J.}\ \bibnamefont
  {Schreier}}\ and\ \bibinfo {author} {\bibfnamefont {L.~L.}\ \bibnamefont
  {Scharf}},\ }\href@noop {} {\emph {\bibinfo {title} {Statistical signal
  processing of complex-valued data: the theory of improper and noncircular
  signals}}}\ (\bibinfo  {publisher} {Cambridge university press},\ \bibinfo
  {year} {2010})\BibitemShut {NoStop}%
\bibitem [{bjo()}]{bjorkman2025observation}%
  \BibitemOpen
  \href@noop {} {\ }\BibitemShut {NoStop}%
\bibitem [{\citenamefont {Kukita}\ \emph
  {et~al.}(2022{\natexlab{a}})\citenamefont {Kukita}, \citenamefont {Kiya},\
  and\ \citenamefont {Kondo}}]{kukita2022general}%
  \BibitemOpen
  \bibfield  {author} {\bibinfo {author} {\bibfnamefont {S.}~\bibnamefont
  {Kukita}}, \bibinfo {author} {\bibfnamefont {H.}~\bibnamefont {Kiya}},\ and\
  \bibinfo {author} {\bibfnamefont {Y.}~\bibnamefont {Kondo}},\ }\bibfield
  {title} {\bibinfo {title} {General off-resonance-error-robust symmetric
  composite pulses with three elementary operations},\ }\href@noop {}
  {\bibfield  {journal} {\bibinfo  {journal} {Physical Review A}\ }\textbf
  {\bibinfo {volume} {106}},\ \bibinfo {pages} {042613} (\bibinfo {year}
  {2022}{\natexlab{a}})}\BibitemShut {NoStop}%
\bibitem [{\citenamefont {Kukita}\ \emph
  {et~al.}(2022{\natexlab{b}})\citenamefont {Kukita}, \citenamefont {Kiya},\
  and\ \citenamefont {Kondo}}]{kukita2022short}%
  \BibitemOpen
  \bibfield  {author} {\bibinfo {author} {\bibfnamefont {S.}~\bibnamefont
  {Kukita}}, \bibinfo {author} {\bibfnamefont {H.}~\bibnamefont {Kiya}},\ and\
  \bibinfo {author} {\bibfnamefont {Y.}~\bibnamefont {Kondo}},\ }\bibfield
  {title} {\bibinfo {title} {Short composite quantum gate robust against two
  common systematic errors},\ }\href@noop {} {\bibfield  {journal} {\bibinfo
  {journal} {Journal of the Physical Society of Japan}\ }\textbf {\bibinfo
  {volume} {91}},\ \bibinfo {pages} {104001} (\bibinfo {year}
  {2022}{\natexlab{b}})}\BibitemShut {NoStop}%
\end{thebibliography}%
	
	\clearpage

	\FloatBarrier
	
	\appendix

\section{Experimental setup}\label{exp_setup}
The transmon qubit used in the experiment is a subset of a two-qubit device similar to the quantum processor studied in Ref.~\cite{PhysRevApplied.14.034010}, with two fixed-frequency transmons capacitively coupled via a frequency-tunable coupler. The chip is packaged, protected from the spurious magnetic noise by a $\mu \text{-metal}$ shield and thermally anchored to the mixing chamber of the refrigerator. During the experiments, only one qubit ($Q_1$) is used, and the coupler is detuned to its sweet spot at the maximum frequency to minimize its impact on the qubit.

\begin{figure}
		\centering
		\includegraphics[width=\defaultfigurewidth]{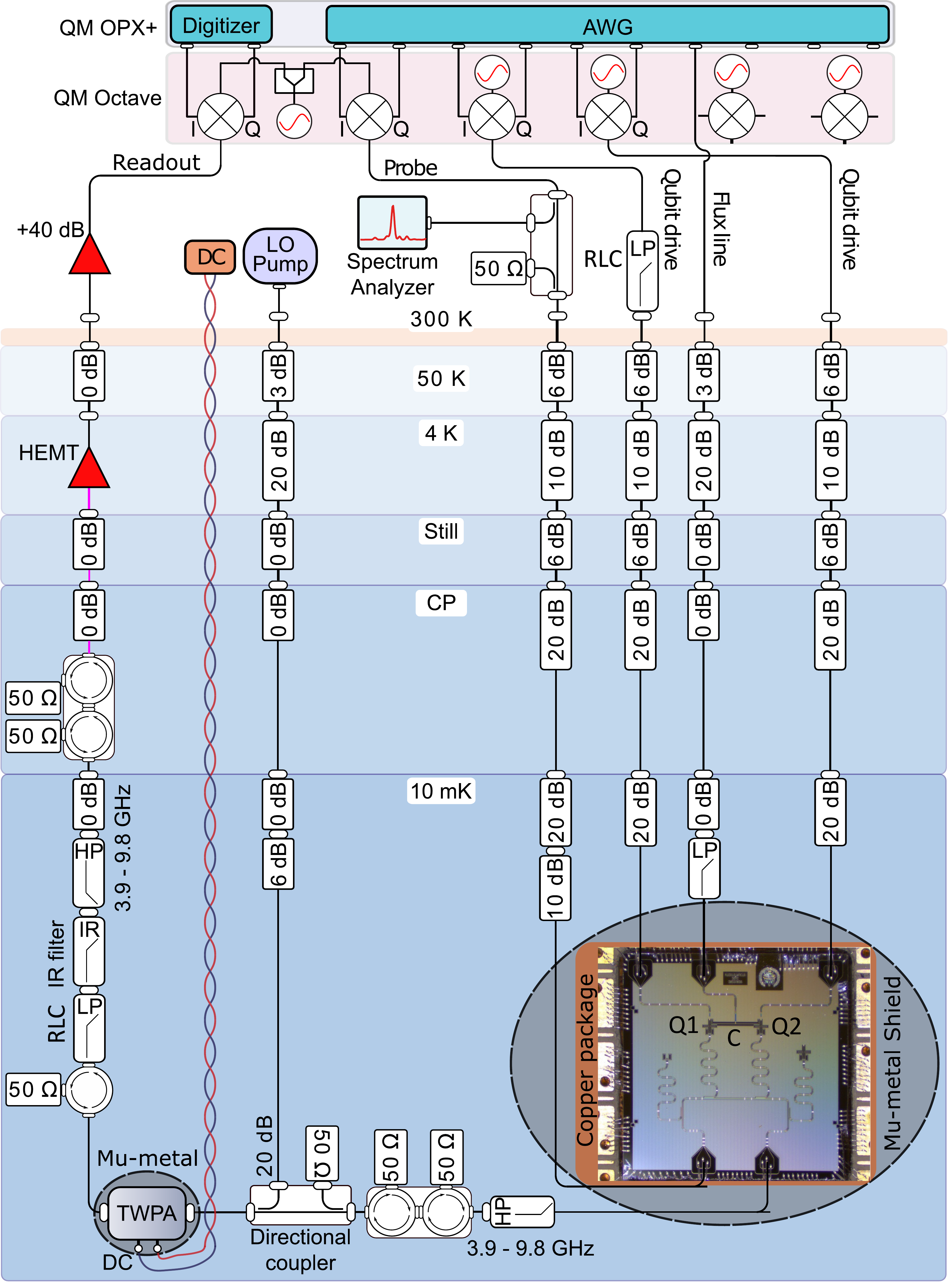}
		\caption{A diagram of the experimental setup. Microscope image of the quantum processor with two transmon qubits. Only the left qubit ($Q_1$) is used for the experiment. The sample is mounted inside a $\mu \text{-metal}$ shield to protect it from stray magnetic fields.
		}
		\label{fig:exp_setup}
	\end{figure}

The experimental setup used is presented in Fig.~\ref{fig:exp_setup}. The pulses are generated at room temperature by using an ensemble OPX+/Octave from Quantum Machines, capable of generating microwave pulses within a 350 MHz bandwidth. To achieve low enough temperatures, we employ a BlueFors XLD400 dilution fridge with base temperature of $\approx10\mathrm{mK}$. Single-shot measurements are enabled by the traveling wave parametric amplifier (TWPA) from VTT Technical Research Center of Finland, while at 4K the readout signal is further amplified by a high-electron-mobility transistor (HEMT) fabricated by Low Noise Factory. After further room-temperature amplification (by Narda Miteq LNA-40-04000800-07-10P), the readout signal is fed back into the OPX+/Octave ensemble, where it is downconverted and digitized.

The pulses used for state preparation and quantum process tomography (cf. section~\ref{QPT}) use a cosine envelope with a $40\mathrm{ns}$ duration combined with the DRAG technique and calibrated according to "Google" method described in~\cite{PhysRevLett.116.020501}. The average gate fidelity characterized with Clifford RB~\cite{PhysRevLett.106.180504} exceeds $99.96\%$.

The singleshot readout uses a $2\mathrm{\mu s}$ pulse, with an average two-state discrimination fidelity of $F_{Assign({|0\rangle,|1\rangle})} = 95.65\%$, and $F_{Assign({|0\rangle,|1\rangle, |2\rangle})} = 94.13\%$ for three states.

	\section{Implementation verification} \label{impl-ver}
	
\begin{figure}
		\centering
		\includegraphics[width=\defaultfigurewidth]{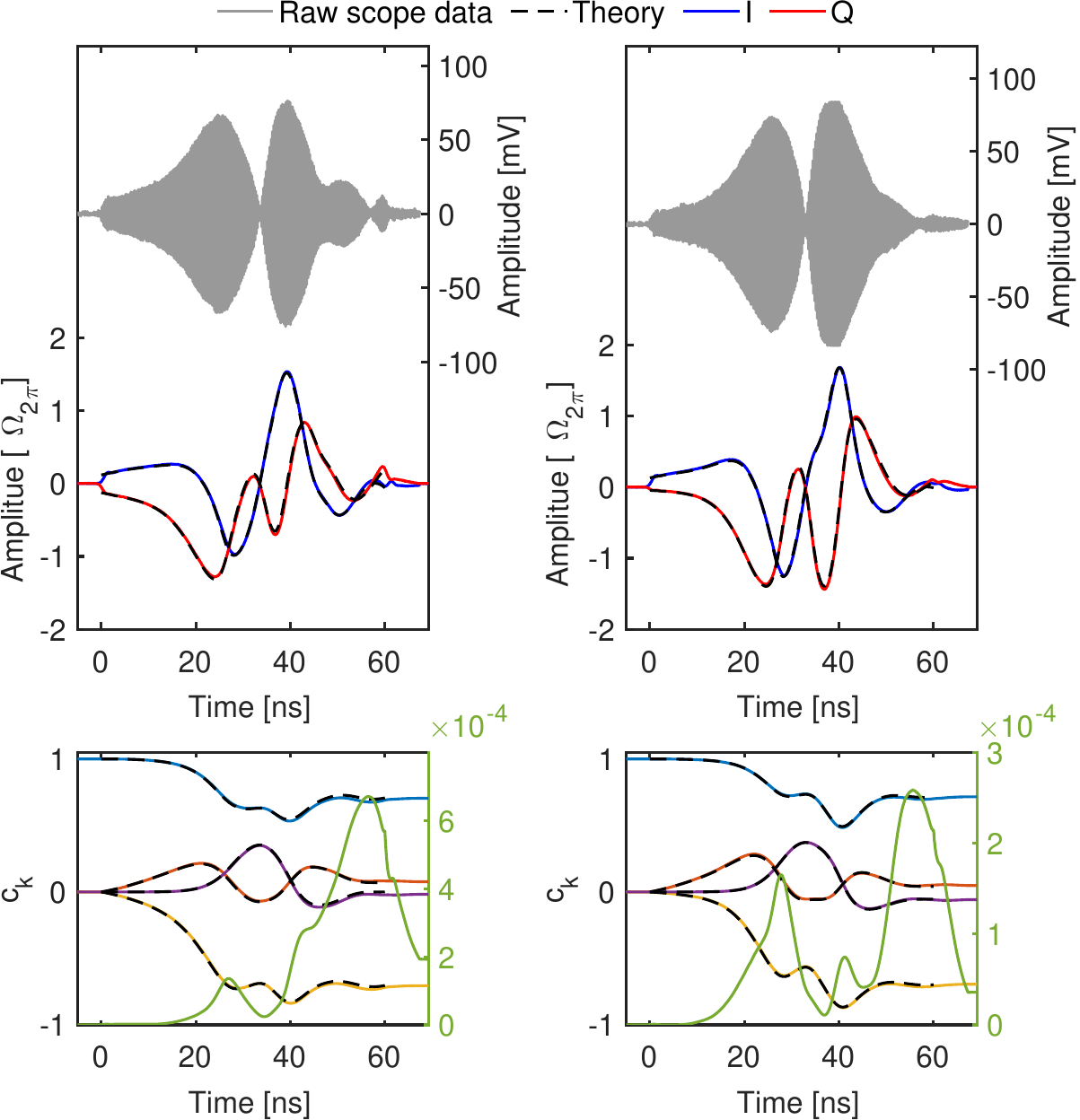}
		\caption{\textbf{Top}: The real and imaginary part of the pulse $\Omega(t) e^{i\varphi(t)}$ (black dashed), along with the measured pulse (gray) and its demodulated real (blue) and imaginary (red) components.
			\textbf{Bottom}: The $c_0$ (blue), $c_x$ (red), $c_y$ (yellow) and $c_z$ (purple) coefficients as a function of time for the scope data, along with the theory (black dashed). The right-hand scale, along with the solid green trace, shows the in-fidelity (1-$F$) of the experimental trajectory with respect to the theoretical one. The \textbf{left} and \textbf{right} columns correspond to the qubit and qutrit optimized pulses, respectively.
		}
		\label{fig:scope_pulses}
	\end{figure}

	The pulses generated by the procedure described in the main text can then be digitized using a sufficiently fast oscilloscope directly (Keysight MSOS804A was used here). With the captured trace (cf. Fig-~\ref{fig:scope_pulses}), an \emph{analytic}, complex valued, signal representation can be obtained by applying a Hilbert transform~\cite{schreier2010statistical}. The analytic form can be easily demodulated by multiplication by $e^{-i \omega_{\rm ge} t}$, with $\omega_{\rm ge} = 2\pi f_{\rm ge}$ being the qubit / signal center frequency. Afterwards it lends itself to comparison to the theory directly, through the real and the imaginary part.
	
	Based on these reconstructed drive amplitudes, the corresponding trajectory can be calculated, which are also shown in Fig.~\ref{fig:scope_pulses}. The fidelity of the reconstructed trajectory, defined as $F=\mathrm{tr}\{U_{\rm exp}(t)U^\dagger_{\rm theory}(t)\}/2$, remains high ($F> 0.999$) throughout the evolution (also in Fig.~\ref{fig:scope_pulses}).

	\section{Quantum process tomography} \label{QPT}

\begin{figure}
		\centering
		\includegraphics[width=\defaultfigurewidth]{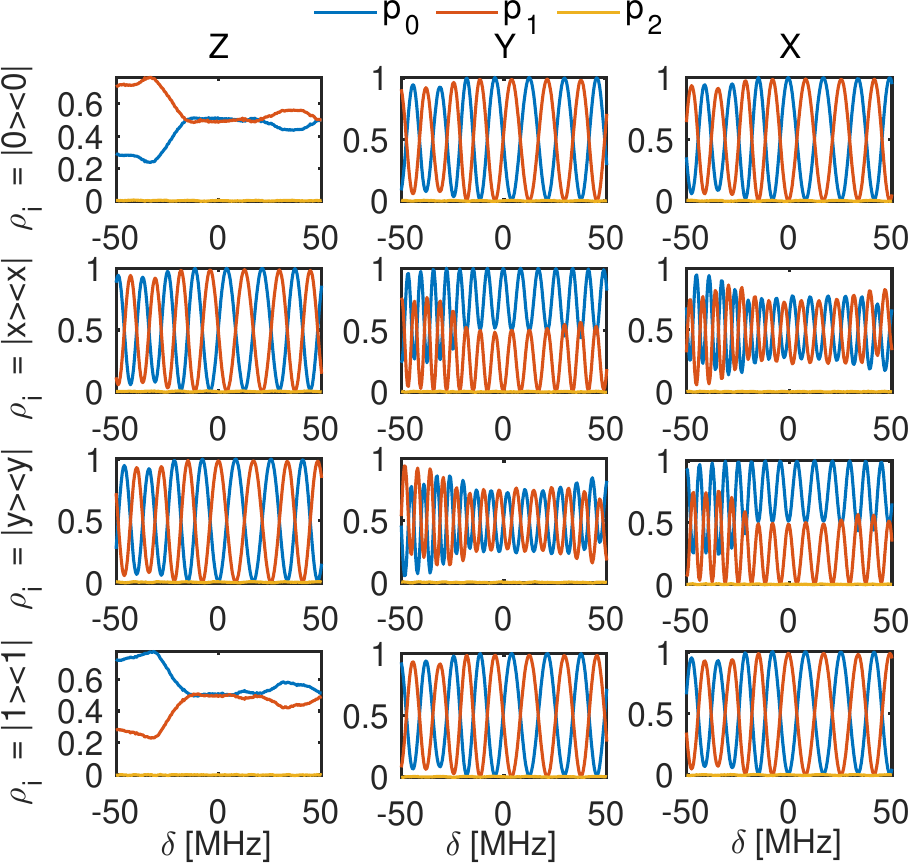}
		\caption{The raw results of the frequency dependent process tomography or the robust $\big(\frac{\pi}{2}\big)$ pulse. Different rows correspond to different initial states, while the columns correspond to reading out the $x$, $y$ and $z$ components of the density matrix. See the text for more details.}
		\label{fig:proc_tomo_states}
	\end{figure}

	Quantum process tomography is performed in the manner described in~\cite{nielsen_chuang_2010}: 4 input states are prepared by initializing the system in the ground state and then applying an identity (no-op) gate, a $(-\frac{\pi}{2})_{X}$, $(-\frac{\pi}{2})_{Y}$, or a $(\pi)_{X}$ gate, resulting in the following initial states $\ket{0}\bra{0}$, $\ket{x}\bra{x}$, $\ket{y}\bra{y}$ and $\ket{1}\bra{1}$ respectively. Then the operation-under-test is applied, followed up by state tomography where the $z$ component is read out directly and for the $y$ and $x$ ones the readout is preceded by applying $(\frac{\pi}{2})_{X}$, $(\frac{\pi}{2})_{Y}$.  This is then performed as a function of the test-pulse frequency; the results for the robust, neural network based, $\big(\frac{\pi}{2}\big)$ are shown in Fig.~\ref{fig:proc_tomo_states}.
	
	From there we see that the population of the $\ket{2}$ state is negligible, allowing us to carry out the process tomography only in the qubit subspace. The detuned pulses are generated by briefly changing the frequency of the IF oscillator in the Quantum Machines OPX+ controller, and the phase difference accumulated during this frequency excursion leads to the rapidly oscillating behavior observed in most panels of Fig.~\ref{fig:proc_tomo_states}, which can later be accounted for.
	
\begin{figure}
		\centering
		\includegraphics[width=\defaultfigurewidth]{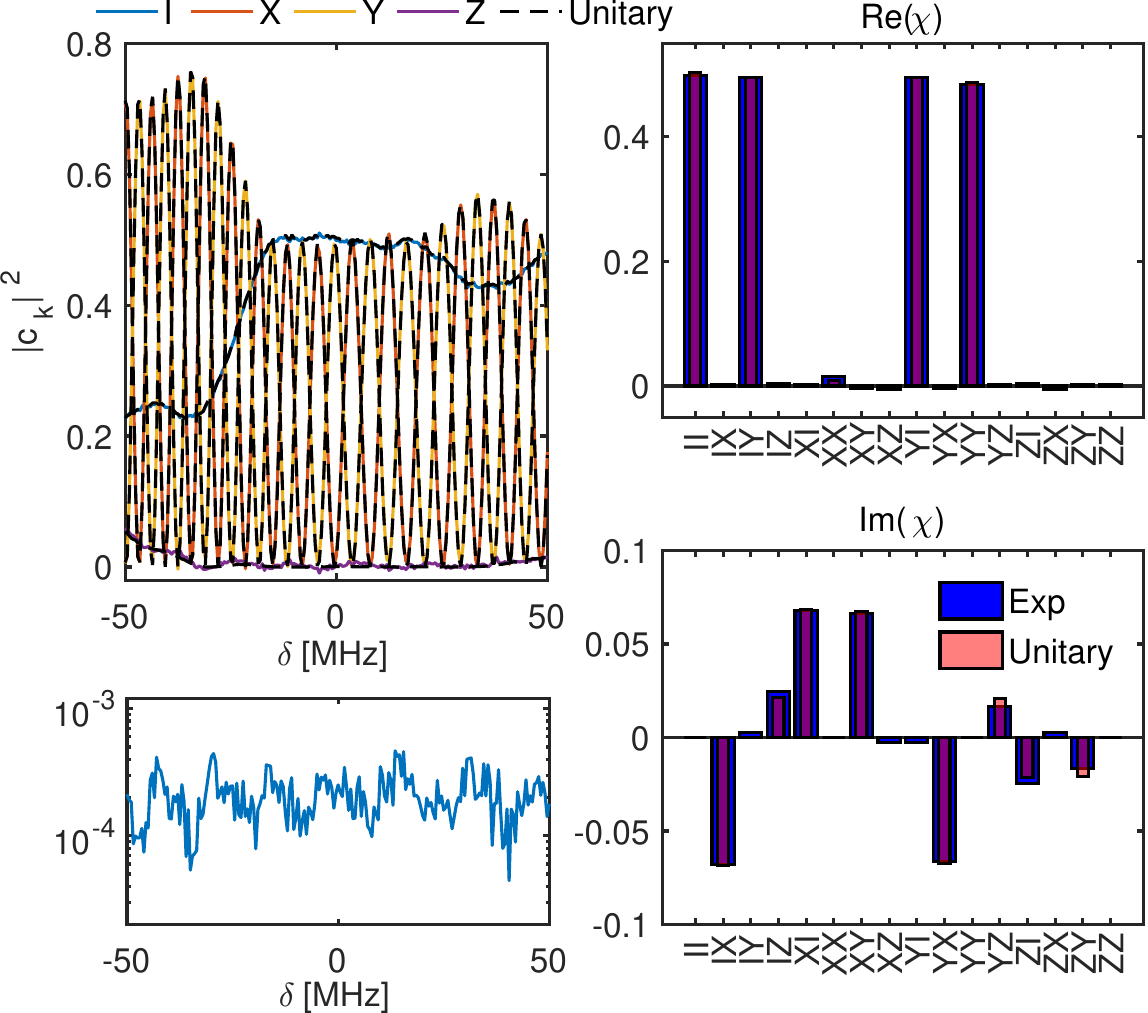}
		\caption{\textbf{Left}: The diagonal elements of the experimental process matrix (solid, colored), along with that of the unitary fit (black, dashed), as a function of detuning $\delta$, along with their Frobenius distance (bottom) . \textbf{Right}: The real (top) and imaginary (bottom) parts of the experimental and unitary process matrices at $\delta=0$.}
		\label{fig:proc_matrix}
	\end{figure}

	The data can be processed to extract the process matrix~\cite{nielsen_chuang_2010}, resulting in a complex $4\times4$ $\chi_{\rm exp}$ matrix, as shown in Fig. ~\ref{fig:proc_matrix}. This experimentally obtained process matrix contains contributions not only from the unitary evolution, but also from relaxation, decoherence and SPAM errors~\cite{PhysRevLett.109.080505}.  For estimating the underlying unitary operation, we fit the experimental process matrix with a unitary one $\chi_{\rm fit}$, by minimizing their Frobenius distance. To gauge the errors introduced by this process we note the following: for a strictly unitary process,  $U = c_0 \sigma_0 - i \sum_{k\in{x,y,z}} c_k \sigma_k$, the diagonal elements of $\chi$ are just $\chi_{k,k} = |c_k|^2$. In Fig.~\ref{fig:proc_matrix} we compare the diagonal elements of $\chi_{\rm exp}$ and that of $\chi_{\rm fit}$, along with their Frobenius distance. As their agreement is quite good, and the distance small, we assume that this leads to a reliable estimate of the implemented unitary evolution, which might not be true for higher SPAM or decoherence errors.
	
	The $c_k$ coefficients shown in figure~\ref{fig:proc_matrix} are identical to the ones presented in fig.~\ref{fig:new_pulse_proc_tomo} in the main text up to an additional rotation in the $X/Y$ subspace.
	As mentioned before, this rotation angle is $\beta = 2 \pi \delta \big(T_{\rm pulse} + T_{\rm dead}\big)$, where $T_{\rm dead}\approx 10~\mathrm{ns}$ is the time needed for the IF oscillator frequency change. Applying the inverse rotation in the $X/Y$ subspace $c'_x + i c'_y = \big(c_x + i c_y\big) e^{-i \beta}$ yields the results presented in fig.~\ref{fig:new_pulse_proc_tomo}.

	\section{Comparison with other pulses}\label{pulse_comparison}
	
	Here we first provide further details about the comparison with the DRAG and rectangular drives. In section~\ref{comparison_composite} we discuss the alternative approach of using composite pulses.
	
	\subsection{DRAG and rectangular drive}\label{comparison_qpt}
		
\begin{figure}
		\centering
		\includegraphics[width=\defaultfigurewidth]{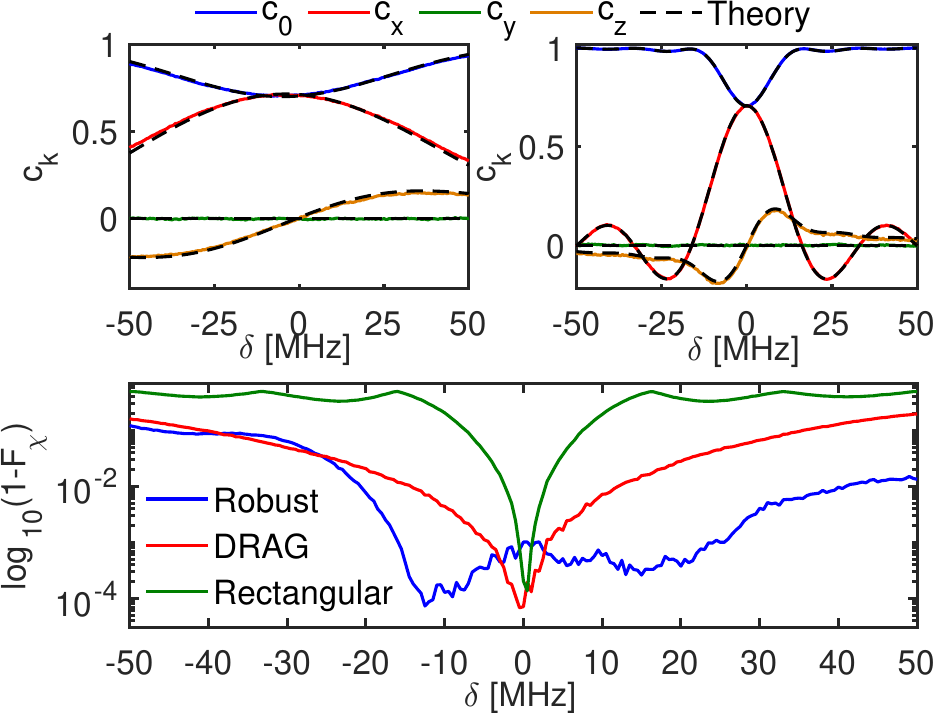}
		\caption{\textbf{Top}: The $c_k$ decomposition of the DRAG (left) and rectangular (right) pulses as a function of detuning. \textbf{Bottom}: the process matrix fidelity $F_\chi$ between the experimentally obtained ones and a vertical $\pi/2$ rotation.}
		\label{fig:drag20_rect60_QPT}
	\end{figure}

	As stated in the main text, the same QPT procedure was carried out for the $T=60~\mathrm{ns}$ rectangular and $T=20~\mathrm{ns}$ pulses, the results of which are shown in Fig.~\ref{fig:drag20_rect60_QPT}.	
	Fig.~\ref{fig:drag20_rect60_QPT} also shows the discrepancy of the implemented operation with respect to a $\big(\frac{\pi}{2}\big)$ rotation in an \emph{arbitrary}, but vertical plane. It is quantified as $F_\chi=\mathrm{tr}(\tilde{\chi}_{\rm exp} \chi_{\pi/2})$, where $\tilde{\chi}_{\rm exp}$ is the \emph{purified} process matrix $\tilde{\chi}_{\rm exp} = \chi_{\rm exp} / \sqrt{\mathrm{tr}(\chi_{\rm exp}^2)}$ and $\chi_{\pi/2}$ is the process matrix corresponding to a unitary with the following expansion $c_0 = 1/\sqrt{2}$, $c_x = \cos(\Theta)/\sqrt{2}$, $c_y = \sin(\Theta)/\sqrt{2}$ and $c_z=0$, where $\Theta$ is chosen such to maximize $F_\chi$.
	The wide range of detuning frequencies for which $F_\chi\approx1$ demonstrates yet again the robustness of the neural network based pulse.

	Much like in Fig.~\ref{fig:p1max_RB} of the main text, from Fig.~\ref{fig:drag20_rect60_QPT} we see that the response of the DRAG pulse is shifted to the left. As mentioned previously this is shift is necessary to cancel the AC Stark effect from the $\ket{2}$ state. Unlike for the robust pulse, where the $\sigma_z$ component $c_z\approx0$ (see fig.~\ref{fig:new_pulse_proc_tomo} of the main text), for here that is not the case. In fact, precisely because of the shift towards $\delta<0$, and cannot be corrected for by an additional detuning.
	
\begin{figure}
		\centering
		\includegraphics[width=\defaultfigurewidth]{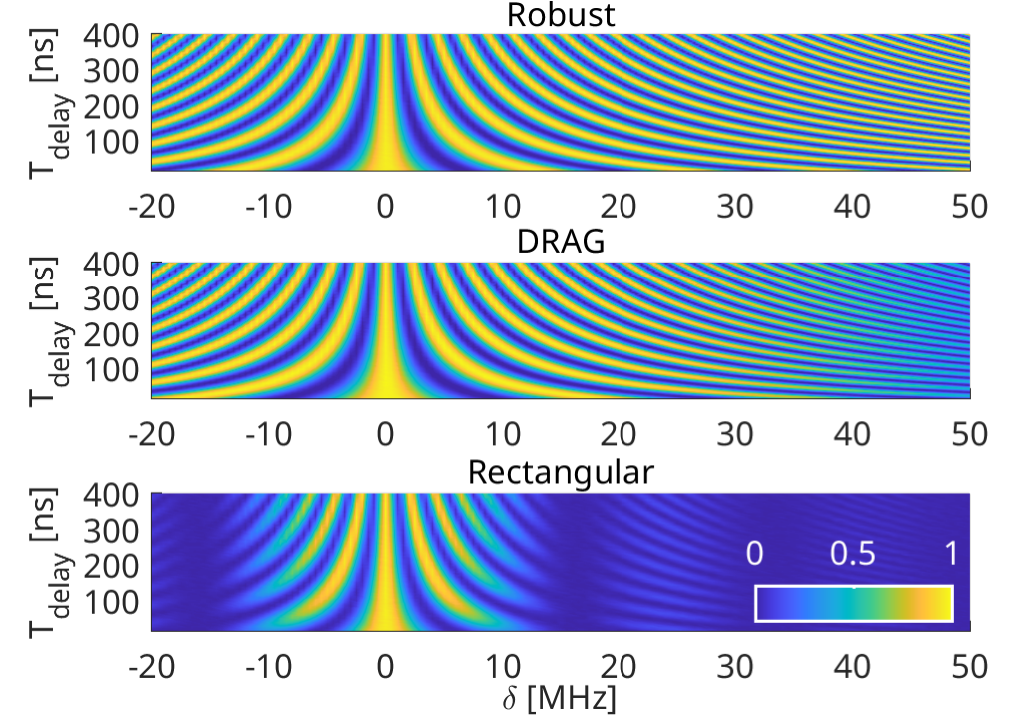}
		\vspace{0mm}
		\includegraphics[width=\defaultfigurewidth]{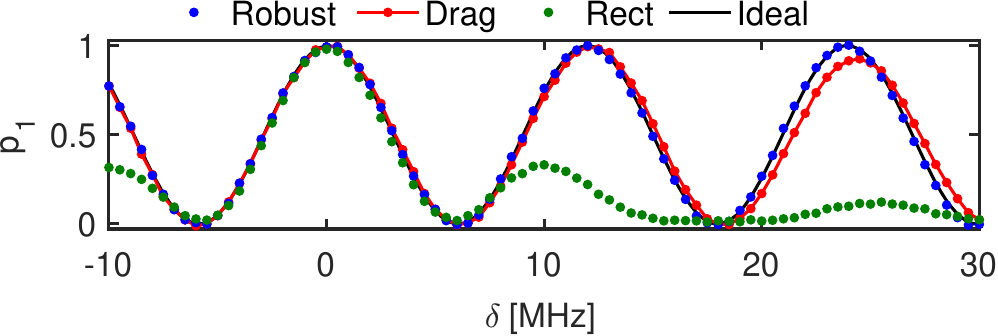}
		\caption{\textbf{Top}: The Ramsey interference pattern measured for the robust, DRAG and rectangular pulses. \textbf{Bottom}: slices of the former as a function of detuning $\delta$ for a fixed $T_{delay}$ (colored symbols), along with the ideal response (solid black).}
		\label{fig:ramsey}
	\end{figure}

	These nonzero $c_z$ components limit the usefulness of DRAG pulses for the parity measurement protocol: Ramsey interference patterns are shown in Fig.~\ref{fig:ramsey} for all three pulses. By taking a slice at a constant protocol time $T_{\rm protocol} = T_{\rm delay} + T_{\rm pulse}$ and plotting the excited state population as a function of frequency $\delta$, also shown in Fig.~\ref{fig:ramsey}, we can see the effect of the $z$ rotation. While the envelope of the oscillation corresponds to the one obtained from measuring $\mathrm{max}(p_1)$ and shown in Fig. ~\ref{fig:p1max_RB} of the main text, we see that the positions of the maxima / minima start deviating from an ideal sinusoidal pattern for $\delta\gg 0$. As a result, the parity measurement implemented with DRAG pulses is limited to lower photon numbers compared to the robust pulse.
	
	To understand this further we can study the following: the idle time $T_{\rm delay}$ is equivalent to a $z$ frame rotation $e^{i \beta \sigma_z /2}$ by the angle $\beta = 2 \pi \delta T_{\rm delay}$. Sandwiching this between two operations, given as $U = c_0 \sigma_0 - i \sum_{k\in{x,y,z}} c_k \sigma_k$, results in the total operation $U_{\rm tot} = U e^{i \beta \sigma_z /2} U$. The squared norm of the matrix element $U_{\rm tot,1,2}=\bra{1}U_{\rm tot}\ket{0}$ is the probability for the transition to occur from the ground to the excited state, which is precisely what we measure experimentally. After some straightforward algebra it evaluates to $|U_{\rm tot,1,2}|^2 = 4 \left(c_x^2+c_y^2\right) \left(c_0 \cos \left(\beta/2 \right)+c_z \sin \left(\beta /2\right)\right)^2 = 4 \left(c_x^2+c_y^2\right) \left(c_z^2+c_x^2\right)\cos \left(\frac{\beta -\beta_0 }{2}\right)^2$, where $\beta_0 = 2\arctan\left(c_z/c_0\right)$. The transition probability reaches unity only when $\left(c_x^2+c_y^2\right) = \left(c_z^2+c_x^2\right) = \frac{1}{2}$ and $\beta  = \beta_0$. However as $\beta_0$ depends non-linearly on $\delta$ through $c_{0,z}(\delta)$, and $\beta$ is a linear function of $\delta$, the last condition cannot be satisfied for all $\delta$ by an appropriate choice of $T_{\rm delay}$.
	
	\subsection{$\max(p_1)$ experiment}
	
\begin{figure}
		\centering
		\includegraphics[width=\defaultfigurewidth]{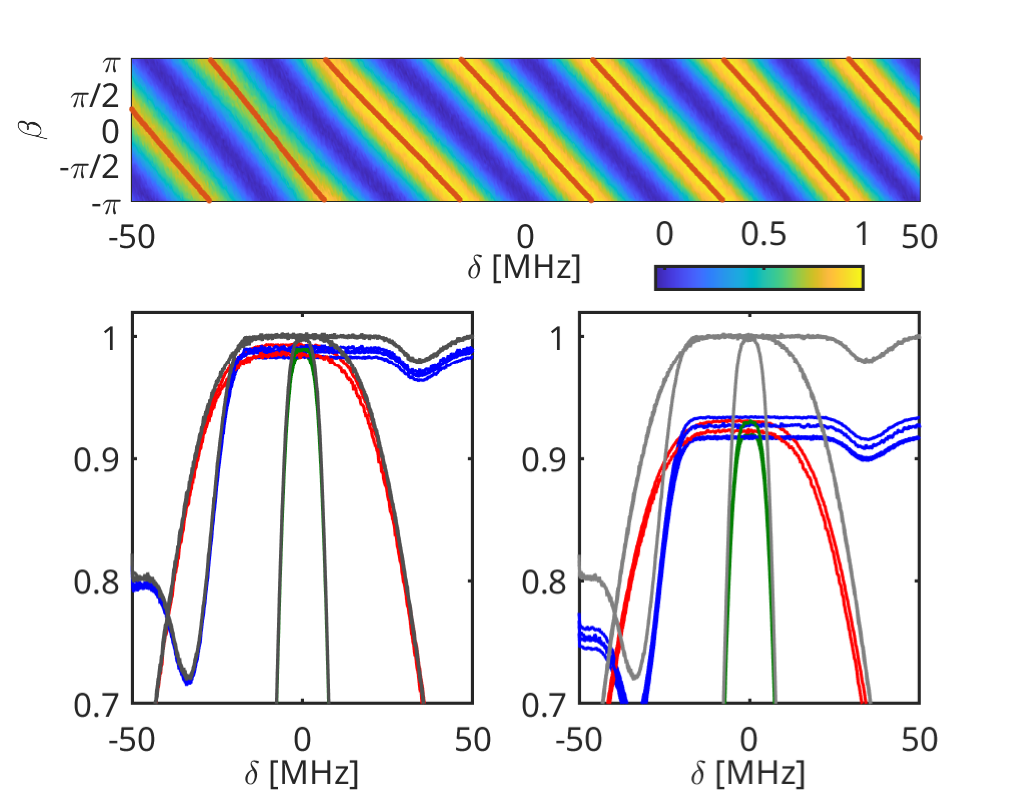}
		\caption{\textbf{Top}: A 2D map of the excited state population as a function of the detuning $\delta$ and relative pulse phase $\beta$. The red dots indicate the optimal $\beta$ values for each $\delta$. \textbf{Bottom}: The measured excited state population as a function of $\delta$, using the optimal $\beta$ values, for the robust (blue), DRAG (red) and rectangular (green) pulses before, and after correcting for the readout fidelity (gray). The left and right panels correspond to the averaged and single-shot readout schemes respectively.}
		\label{fig:p1maxapopcorr2}
	\end{figure}

	As described in section~\ref{comparison_qpt}, the transition $\ket{0} \rightarrow \ket{1}$ probability in a Ramsey-like experiment can be estimated from the QPT data, or theoretical model, as $\max(p_1) = 4 \left(c_x^2+c_y^2\right) \left(c_z^2+c_x^2\right)$, which is obtained for an accumulated phase $\beta = k 2\pi + \beta_0$. This implies that the delay time in the Ramsey sequence cannot be chosen solely based on the detuning; it requires prior knowledge of the $c_{0,z}(\delta)$ dependence. Hence we go about measuring $\max(p_1)$ by applying a \emph{virtual} $z$ rotation in-between the two drive pulses. A 2D map which measures the transition probability, as a function of detuning $\delta$ and the frame rotation angle $\beta$, is shown in Fig.~\ref{fig:p1maxapopcorr2}. As the $\beta$ dependence is sinusoidal the optimal value can easily be determined by performing a Fourier-like fit. 
	
	Using these optimal values, we proceed to measure the excited state population as a function of $\delta$. The state of the transmon is read-out by probing a dispersively coupled resonator, either in a single-shot fashion or in the averaged regime. The first is foremost limited by the SNR of the readout and the system relaxation times relative to the readout duration, while the latter is sensitive mostly to the post-processing procedure used to extract the populations, as described in detail in~\cite{bjorkman2025observation}. As we need to resolve minute variations in $p_1$ a lot of measurement shots are required to discern them; we used $N=5\times 10^5$ shots. While this results in very low shot-noise for the singleshot and good SNR for the averaged readout scheme, the long measurement time makes the experiment highly susceptible to errors induced by drifts in the amplification / readout chain. Figure~\ref{fig:p1maxapopcorr2} shows several datasets, for each of the studied pulses, where singleshot and averaged readout were performed simultaneously. Both readout schemes suffer from variations and drifts in their efficiency, which amounts to a systematic error present in the readout procedure.	
	However, when one accounts for the variations, all of the datasets collapse to the same trace. This, along with the excellent agreement with the QPT-based estimate, gives validity to the correction procedure.

	\subsection{Composite pulses}\label{comparison_composite}

	While making an exhaustive comparison with other optimal control procedures is not feasible, we decide to compare our scheme with a short composite pulse sequence. The rationale for this is the following: the robust pulse is comparatively long $T=60~\mathrm{ns}$ to the typical durations of DRAG pulses $T=\numrange{10}{20}~\mathrm{ns}$. Thus a sequence of 3 to 6 pulses can be implemented in the same duration.
	
\begin{figure}
		\centering
		\includegraphics[width=\defaultfigurewidth]{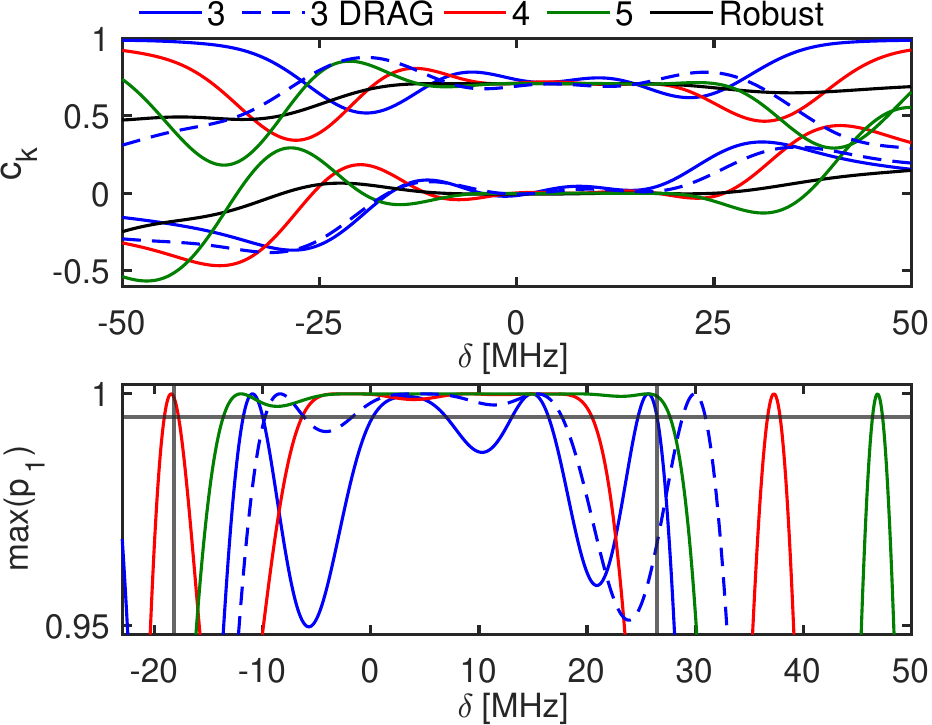}
		\caption{\textbf{Top}: The $c_0$ and $c_z$ traces for various composite-pulse sequence lengths. The solid-colored lines correspond to the rectangular pulses, the black ones to the robust pulse, and the dashed-blue to a 3-pulse long sequence implemented with DRAG pulses. \textbf{Bottom}: The corresponding transition probabilities in a Ramsey-like experiment; the solid gray lines indicate the region in which the robust pulse has an efficiency greater than $0.995$.}
		\label{fig:corpse}
	\end{figure}

	Famously the CORPSE~\cite{cummins2000use} sequence even provides an analytical recipe how to implement a 3-pulse long detuning-robust drive. These results were later generalized by e.g.,~\cite{kukita2022general}, and longer sequences have also been studied~\cite{kukita2022short}. However, they rely on first- or second- order error cancellation around $\delta=0$, which does not yield good performance for larger detuning values, which are of relevance here. 
	Thus, instead of relying on analytical models we numerically searched for the optimal sequences, 3 to 5 pulses long, using a combination of a genetic algorithm (for the sake of parameter-space exploration) and refining the result using a gradient-based optimizer. The loss function to be optimized was made identical to the one used for the NN based approach. For the sake of visual clarity, in figure~\ref{fig:corpse} we report only the $c_0$ and $c_z$ components of the optimal composite pulses, as well as the the corresponding $\max(p_1)$ traces. We find that for both metrics the composite sequences exhibit larger variations than the NN pulse, which is to be expected due to the low number of parameters. On the other hand, these numerically obtained sequences vastly outperform the analytical solutions (not shown).
	
	The claim, made in~\cite{cummins2000use}, that the spectral content of the pulses does not affect the performance of the composite sequence is true near $\delta=0$: close to resonance the DRAG and the rectangular pulse look the same to first order, if made the same duration - see Fig.~\ref{fig:drag20_rect60_QPT}. However, for larger $|\delta |$ the argument breaks down. When attempting to implement a 3-pulse long sequence using DRAG pulses, the spectral response changes substantially leading to a different result, even after optimization - see Fig. ~\ref{fig:corpse}. This finding further complicates implementing a composite pulse sequence based on DRAG: the DRAG coefficient is not rotation-angle independent, and the spectral response is sensitive to it's value. Thus, to properly fine-tune the sequence one must perform the optimization in a closed loop, where the sequence and the pulses themselves are tuned simultaneously, making it costly both numerically and experimentally.
	
\end{document}